\documentclass[11pt]{article}

\usepackage{graphicx}
\graphicspath{{figures/}}
\usepackage{mathtools}
\usepackage{amssymb}
\usepackage{xcolor}
\usepackage{subfigure}
\usepackage{caption}

\usepackage[font=footnotesize,labelfont=bf]{caption}

\usepackage[T1]{fontenc}

\newcommand{\norm}[1]{\left\lVert#1\right\rVert}

\usepackage{fancyhdr}
\fancypagestyle{firstpage}{\fancyhf{}\fancyhead[R]\fancyfoot{New York, \today}}
\fancypagestyle{plain}{\fancyhf{} \fancyhead[R]{\thepage}}

\usepackage{geometry}
\long\def\comment#1{}

\definecolor{darkblue}{rgb}{0.1,0,0.6}
\definecolor{green}{rgb}{0,1,0}

\begin{document}

\section*{\LARGE Neural tuning and representational geometry}

Nikolaus Kriegeskorte\textsuperscript{1,2,3~*~$\dagger$} and Xue-Xin Wei\textsuperscript{4,5,6,7~*~$\dagger$}

1) Zuckerman Mind Brain Behavior Institute, Columbia University, New York, NY, USA \\
2) Department of Psychology, Columbia University, New York, NY, USA \\
3) Department of Neuroscience, Columbia University, New York, NY, USA \\
4) Department of Neuroscience, University of Texas at Austin, Austin, TX, USA\\
5) Department of Psychology, University of Texas at Austin, TX, USA\\
6) Center for Perceptual Systems, University of Texas at Austin, TX, USA\\
7) Institute for Neuroscience, University of Texas at Austin, TX, USA

\medskip
* = co-first authors;  $\dagger$ = co-senior authors. Listed alphabetically.

\medskip
Correspondence: nk2765@columbia.edu; weixx@utexas.edu

\section*{Abstract}

A central goal of neuroscience is to understand the representations formed by brain activity patterns and their connection to behavior. The classical approach is to investigate how individual neurons encode the stimuli and how their tuning determines the fidelity of the neural representation. Tuning analyses often use the Fisher information to characterize the sensitivity of neural responses to small changes of the stimulus. In recent decades, measurements of large populations of neurons have motivated a complementary approach, which focuses on the information available to linear decoders. The decodable information is captured by the geometry of the representational patterns in the multivariate response space. Here we review neural tuning and representational geometry with the goal of clarifying the relationship between them. The tuning induces the geometry, but different sets of tuned neurons can induce the same geometry. The geometry determines the Fisher information, the mutual information, and the behavioral performance of an ideal observer in a range of psychophysical tasks. We argue that future studies can benefit from considering both tuning and geometry to understand neural codes and reveal the connections between stimulus, brain activity, and behavior.

\newpage
\pagestyle{plain}

\section{Introduction}

Brain computation can be understood as the transformation of representations across stages of processing \cite{decharms2000neural}. Information about the world enters through the senses, is encoded and successively re-coded so as to extract behaviorally relevant information
and transform its representational format to enable the inferences and decisions needed for successful action. Neuroscientists often focus on the code in one particular neural population at a time (let's call it area X), summarily referring to the processes producing the code in X as the encoder, and the processes exploiting the code in X as the decoder~\cite{kriegeskorte2019interpreting}.

The neural code in our area of interest X is classically characterized in terms of the \emph{tuning} of individual neurons. A tuning curve describes the dependence of a neuron's firing on a particular variable thought to be represented by the brain. Like a radio tuned to a particular frequency to select a station, a "tuned" neuron~\cite{Barlow1967} may selectively respond to stimuli within a particular band of some stimulus variable, which could be a spatial or temporal frequency, or some other property of the stimulus such as its orientation, position, or depth~\cite{Barlow1967,Campbell1968,Blakemore1972,Henry1974}. As a classic example, the work by Hubel and Wiesel~\cite{Hubel1959,Hubel1962,Hubel1968} demonstrated that the firing rate of many V1 neurons is systematically modulated by the retinal position and orientation of an edge visually presented to a cat. The dependence was well described by a bell-shaped tuning curve~\cite{Campbell1968,Blakemore1972,Rose1974,Henry1974,Swindale1998}. Tuning provides a metaphor and the tuning curve a quantitative description of a neuron's response as a function of a particular stimulus property. A set of tuning curves can be used to define a neural population code~\cite{Georgopoulos1986,Ben1995,Anderson2000,Series2004,Butts2006}.
Consider the problem of encoding a circular variable, such as the orientation of an edge segment, with a population of neurons. A canonical model assumes that each neuron has a (circular) Gaussian tuning curve with a different preferred orientation. The preferred orientations are distributed over the entire cycle, so the neural population can collectively encode any orientation.

Over the past several decades, systems neuroscience has characterized neural tuning in different brain regions and species for a variety of stimulus variables and modalities~(\emph{e.g.,}~\cite{Hubel1962,Campbell1969,Goldberg1969,Suga1977,o1978,Knudsen1982,Maunsell1983,Georgopoulos1986,Taube1990,Deangelis1991,Johnson1992,Gallant1996,Pasupathy2002,Nieder2002,Fyhn2004,Hafting2005,Young1992,Tsao2006,Rigotti2013,Chang2017}). We have also accumulated much knowledge on how tuning changes with spatial context~\cite{Maffei1976,Gilbert1990,Knierim1992}, temporal context ~\cite{Maffei1973,Movshon1979,Dragoi2000,Benucci2013,Dean2005,Ulanovsky2004,Grill2006,Solomon2014,Alink2018}, internal states of the animal such as attention~\cite{Treue1999,McAdams1999,Reynolds2000}, and learning~\cite{Schoups2001,Ghose2002,Crist2001}. 
The tuning curve is a descriptive tool, ideal for characterizing strong dependencies, as found in sensory cortical areas, of individual neurons' responses on environmental variables of behavioral importance. However, when the tuning curve is reified as a computational mechanism, it can detract from non-stimulus-related influences on neuronal activity, such as the spatial and temporal environmental context and the internal network context, where information is encoded in the population activity, and computations are performed in a nonlinear dynamical system  ~\cite{Churchland2012,Shenoy2013} whose trial-to-trial variability may reflect stochastic computations ~\cite{Churchland2011}. Despite their limitations, tuning functions continue to  provide a useful first-order description of how individual neurons \textit{encode} behaviorally relevant information in sensory, cognitive, and motor processes.

A complementary perspective is that of a downstream neuron that receives input from a neural population and must read out particular information toward behaviorally relevant inferences and decisions. This \textit{decoding} perspective addresses the question what information can be read out from a population of neurons using a simple biologically plausible decoder, such as a linear decoder. Decoding analyses have gained momentum more recently, as our ability to record from a large number of channels (\emph{e.g.}, neurons or voxels) has grown with advances in cell recording~\cite{Buzsaki2004,Stevenson2011,Jun2017} and functional imaging~\cite{Biswal1995,Fox2007}. A linear decoding analysis reveals how well a particular variable can be read out by projecting the response pattern onto a particular axis in the population response space. A more general characterization of a neural code is provided by the representational geometry~\cite{Shepard1970,Edelman1998a,Edelman1998b,Norman2006,Diedrichsen2017,Kriegeskorte2008a,Kriegeskorte2008b,Connolly2012,Xue2010,Khaligh2014,Yamins2014,Cichy2014,Freeman2018,Kietzmann2019}. The representational geometry captures all possible projections of the population response patterns and characterizes all the information available to linear and nonlinear decoders ~\cite{Kriegeskorte2019}. The representational geometry is defined by all pairwise distances in the representational space among the response patterns corresponding to a set of stimuli. The distances can be assembled in a representational dissimilarity matrix (RDM), indexed horizontally and vertically by the stimuli, where we use ``dissimilarity" as a more general term that includes estimators that are not distances in the mathematical sense \cite{Kriegeskorte2019}]. The RDM abstracts from the individual neurons, providing a sufficient statistic invariant to permutations of the neurons and, more generally, to rotations of the ensemble of response patterns. These invariances enable us to compare representations among regions, individuals, species and between brains and computational models~\cite{Kriegeskorte2008,Nili2014,Khaligh2014,Kriegeskorte2016}.

The neural tuning determines the representational geometry, so the two approaches are deeply related. Nevertheless they have originally been developed in  disconnected literatures. Tuning analysis has been prevalent in neuronal recording studies, often using simple parametric stimuli and targeting lower-level representations.  The analysis of representational geometry has been prevalent in the context of large-scale brain-activity measurements, especially in functional MRI (fMRI)~\cite{Kriegeskorte2013}, often using natural stimuli and targeting higher-level representations. In recent years, there has been growing cross-pollination between these literatures. Notably, encoding models for simple parametric~\cite{Dumoulin2008} and natural stimuli~\cite{Kay2008,Naselaris2009} have been used to explain single-voxel fMRI responses, and a number of studies have characterized neural recordings in terms of representational geometry~\cite{Kriegeskorte2008b,Freiwald2010,Ringach2019,Khaligh2014,Yamins2014}.

Tuning and geometry are closely related, and the respective communities using these methods have drawn from each other. Here we clarify the mathematical relationship of tuning and geometry and attempt to provide a unifying perspective. We review the literature and explain how tuning and geometry relate to the quantification of information-theoretical quantities such as the Fisher information (FI)~\cite{Fisher1922} and the mutual information (MI)~\cite{Shannon1948}. Examples from the literature illustrate how the dual perspective of tuning and geometry illuminates the content, format, and computational roles of neural codes.

\begin{figure}
\centering
\includegraphics[keepaspectratio,width=.99\linewidth]{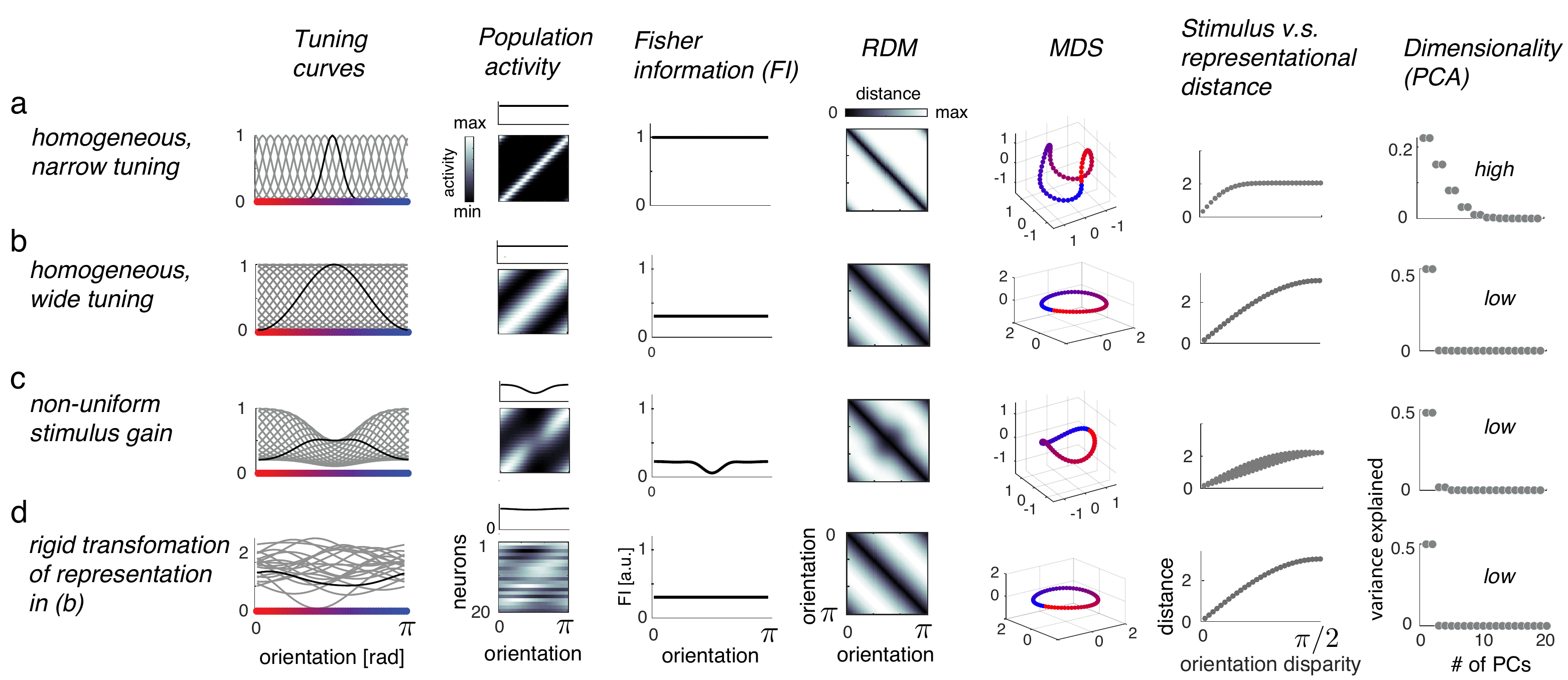}
\caption{\textbf{Neuronal tuning determines representational geometry}. Each row shows a specific configuration of the neural code for encoding a circular variable, such as orientation. Each column shows a particular way to characterize the neural code. The first two columns show traditional tuning-based characterizations. The small inset on top of the population response matrix plots the summed population activity as a function of the stimulus orientation.
The third column plots the Fisher information of the neural population as a function of the stimulus (arbitrary units), which provides a local measure of the coding fidelity. Specifically, the square root of Fisher information gives a lower bound of the discrimination threshold in fine discrimination tasks. The last four columns show the geometry of the neural code, using a representational dissimilarity matrix (RDM), 3-d  multi-dimensional scaling, and the relation between distance in the stimulus space (horizontal axis) and representational distance (vertical axis). The last column plots the proportion of variance explained by individual principal components. This reflects the dimensionality of the representation. Several general insights emerge from this framework:
(1) Tuning width can modulate the representational geometry, without changing the allocation of Fisher information (compare rows \textbf{a} and \textbf{b}).
(2) Fisher information only provides a local measure of the geometry. It does not fully determine the geometry (compare \textbf{a}, \textbf{b}, \textbf{d}).
(3) Non-uniform stimulus gain (\textbf{c}) can change the representational precision (as reflected in the Fisher information) for particular parts of the stimulus space.
(4) The same representation can manifest itself in different neural response patterns (compare \textbf{b} and \textbf{d}). The configuration in (\textbf{d}) is obtained by a rigid transformation (shift and rotation) of the codes in (\textbf{b}).}

\label{fig:setup}
\end{figure}

\noindent\fbox{
\footnotesize{
\parbox{0.98\textwidth}{
        {\bf BOX 1: Mathematical definitions} \\
{\bf \textit{Entropy.}} The entropy of a random variable $\mathbf{r}$ with a probability density $p(\mathbf{r})$ is a measure of our uncertainty about the variable and is defined as:
\vspace{-2mm}
\begin{equation}
    H(\mathbf{r}) = -\int p(\mathbf{r})\cdot \log p(\mathbf{r}) \  d\mathbf{r}
\end{equation}
The conditional entropy of a random variable $\mathbf{r}$ given another random variable $\theta$ is defined as: 
\vspace{-1.5mm}
\begin{equation}
    H(\mathbf{r}|\theta) =-\int{p(\mathbf{r},\theta) \cdot log \ p(\mathbf{r}|\theta) \ d \mathbf{r} \ d \theta}
\end{equation}
{\bf \textit{Mutual information.}} The mutual information (MI) $I(\mathbf{r};\theta)$ between response $\mathbf{r}$ and stimulus $\theta$ measures the information that the response conveys about the stimulus (and vice versa). The MI can be thought of as the shared entropy, or entropy overlap, that is the number of bits by which the entropy $H(\mathbf{r},\theta)$ of the joint distribution of both variables is smaller than the sum of their individual entropies $H(\mathbf{r})$ and $H(\theta)$:
\vspace{-1.5mm}
\begin{equation}
    I(\mathbf{r};\theta)= H(\mathbf{r})+H(\theta) - H(\mathbf{r},\theta)
\end{equation}
Equivalently, the MI is the reduction of the entropy of one of the variables when we learn about the other:
\vspace{-1.5mm}
\begin{equation}
    I(\mathbf{r};\theta)= H(\mathbf{r})-H(\mathbf{r}|\theta) = H(\theta)-H(\theta|\mathbf{r})
\end{equation}
Note that MI depends on the distribution of both variables. In more explicit terms, the MI can be expressed as: 
\vspace{-1mm}
\begin{equation}
     I(\mathbf{r};\theta) = \int \int p(\mathbf{r},\theta)\log \frac{p(\mathbf{r},\theta)}{p(\mathbf{r})p(\theta)}d\mathbf{r}d\theta.
\end{equation}
\comment{
\begin{equation}
     I_{i,j}(\mathbf{r};\{\theta_i,\theta_j\}) = \sum_{k \in \{i,j\}} \int p(\mathbf{r},\theta_k)\log \frac{p(\mathbf{r},\theta_k)}{p(\mathbf{r})p(\theta_k)}d\mathbf{r}.
\end{equation}
}

{\bf \textit{Fisher information.}} The Fisher information (FI) of a neural population code is a function of the stimulus. For each location in stimulus space, the FI reflects how sensitive the code is to local changes of the stimulus. Assume a population of $N$ neurons encoding stimulus variable $\theta$, with tuning curves $f_{i}$. The population response is denoted as $\mathbf{r}$.
The FI can be defined as
\vspace{-2mm}
\begin{equation}
  J(\theta)=\int -\frac{d^2}{d\theta^2} \log \ p(\mathbf{r}| \theta) p(\mathbf{r}) d\mathbf{r}.
\end{equation}         
         
If $\theta$ is a vector, rather than a scalar, description of the stimulus, $ J(\theta)$ becomes the FI matrix, characterizing the magnitude of the response pattern change for all directions in stimulus space. The FI of a neuron is determined by the neuron's tuning curve and noise characteristics. Assuming Poisson spiking, and tuning curve $f(\theta)$,  where $\theta$ is the encoded stimulus property, the Fisher information $J(\theta)$ can be computed as
\vspace{-1.5mm}
\begin{equation}
J(\theta) = \frac{f'(\theta)^2}{f(\theta)}.
\end{equation} Under the assumption of additive stimulus-independent Gaussian noise (with noise variance equal to 1), the Fisher information is simply the squared slope of the tuning curve
\vspace{-2mm}
\begin{equation}
J(\theta) = f'(\theta)^2.
\end{equation} 
For a neural population, if neurons have independent noise, the population FI is simply the sum of the FI for individual neurons. With correlated noise, evaluating the FI is more complicated as it depends on the correlation structure. FI depends on the specific parameterization of the stimulus space. More precisely, when  $\theta$ is reparameterized as $\tilde{\theta}$, the following relation holds:
\vspace{-2mm}
\begin{equation}
\sqrt{J(\theta)}d\theta=\sqrt{J(\tilde{\theta})}d\tilde{\theta}.
\end{equation}{\bf \textit{Discrimination threshold.}} The square root of the Fisher information provides a lower bound on the discrimination threshold $\delta(\theta)$ in fine discrimination tasks
\vspace{-2mm}
\begin{equation}
\delta (\theta) \geq C/\sqrt{J(\theta)},
\end{equation}
where $C$ is a constant dependent on the definition of the psychophysical discrimination threshold.
\vspace{1 mm}\\
{\bf \textit{Variance-stabilizing transformation.}} For independent Poisson noise,  we can apply a square-root transformation to an individual neuron's response $r_i$~\cite{Bartlett1936,Anscombe1948}. The resulting response will have approximately constant variance independent of firing rate. In general, assuming the variance of $r_i$ is a smooth function $g(\cdot)$ of its mean $\mu$ , one could first stabilize the variance by applying the following transformation to obtain $\tilde{r_i}$ and the statements below would apply to $\tilde{r_i}$.
\vspace{-1.5mm}
\begin{equation}
\tilde{r_i}=\int ^{r_i}\frac{1}{\sqrt{g(\mu)}}d\mu
\end{equation}

}}
}

\section{A dual perspective}

\subsection{Tuning determines mutual information and Fisher information}
A tuning curve tells us how the mean activity level of an individual neuron varies depending on a stimulus property. The tuning curve, thus, defines how the stimulus property is \textit{encoded} in the neuron's activity. Conversely, knowing the activity of the neuron may enable us to \textit{decode} the stimulus property. How informative the neural response is depends on the shape of the tuning curve. A flat tuning curve, often described as a lack of tuning, would render the response uninformative. If the neuron is tuned, it contains some information about the stimulus property. However, a single neuron may respond similarly to different stimuli. For example, a neuron tuned to stimulus orientation may respond equally to two stimuli differing  from its preferred orientation in either direction by the same angle. The resulting ambiguity may be resolved by decoding from multiple neurons. 

For a population of neurons, the tuning function $\mathbf{f}(\theta)$ can be defined as the expected value of the response pattern $\mathbf{r}$ given stimulus parameter $\theta$:
\begin{equation}
    \mathbf{f}(\theta) = \mathop{\mathbb{E}_{\mathbf{r} \sim p(\mathbf{r}|\theta)}} [\mathbf{r}]
    = \mathop{\mathbb{E}} [\mathbf{r}|\theta].
\end{equation}

In the special case of additive noise with mean zero, the response pattern obtains as $\mathbf{r} = \mathbf{f}(\theta) + \mathbf{\epsilon}$, where $\mathbf{f}(\theta)$ is the encoding model that defines the tuning function for each neuron and $\mathbf{\epsilon}$ is the noise.

We can quantify the information contained in the neural code about the stimulus using the mutual information or the Fisher information, two measures that have been widely used in computational neuroscience. Mutual information (MI) is a key quantity of information theory that measures how much information one random variable contains about another random variable \cite{Shannon1948}. The MI $I(\mathbf{r};\theta)$ measures the information the response $\mathbf{r}$ conveys about the stimulus $\theta$ (Box 1). Computing the MI requires assuming a prior $p(\theta)$ over the stimuli, because the MI is a function of the joint distribution $p(\mathbf{r},\theta) = p(\mathbf{r}|\theta)\cdot p(\theta)$ of stimulus and response.

Over the past few decades, MI has been used extensively in neuroscience~\cite{van1997,Rieke1999,Borst1999,Fairhall2001}. For example, studies have aimed to quantify the amount of information conveyed by neurons about sensory stimuli \cite{Theunissen1991,van1997,Fairhall2001,Borst1999,Roddey1996,Theunissen1996}.
MI has also been used as an objective function to be maximized, so as to derive an efficient encoding of a stimulus property in a limited number of neurons (\cite{Brenner2000,Ganguli2014,Wei2015,Wei2012,Mcdonnell2008}). If a neural code optimizes the MI, the code must reflect the statistical structure of the environment. For example, in an MI-maximizing code, frequent stimuli should be encoded with greater precision. Once a prior over the stimuli, a functional form of the encoding, and a noise model have been specified, the joint distribution of stimulus and response is defined, and the MI can be computed. The parameters of the encoding can then be optimized to maximize MI~\cite{Barlow1961,Linsker1988,laughlin1981}. If the tuning functions that maximize MI match those observed in brains, the function to convey information about the stimulus may explain why the neurons exhibit their particular tuning~\cite{laughlin1981,van1992,Atick1992,Dong1995,Olshausen1996,Bell1997,Simoncelli2001,Fairhall2001,Ganguli2010,Wei2016}.  The applications of information theory in neuroscience have been the subject of previous reviews (\cite{Rieke1999,Borst1999,Quiroga2009}).

The MI provides a measure of the overall information the response conveys about the stimulus. It can be thought of as the expected value of the reduction of the uncertainty about the stimulus that an ideal observer experiences when given the response, where the expectation is across the distribution of stimuli. However, the MI does not capture how the fidelity of the encoding depends on the stimulus. Some authors have proposed stimulus-specific variants of MI~\cite{deweese1999,Butts2003,Butts2006,Montgomery2010}.

Another stimulus-specific measure of information in the neural response is the Fisher information (FI). The FI has deep roots in statistical estimation theory~\cite{Fisher1922,Lehmann2006} and plays a central role in the framework we propose here. The FI captures the precision with which a stimulus property is encoded in the vicinity of a particular value of the property~\cite{Harper2004,Gutnisky2008,Brunel1998,Ganguli2014,Wei2016}. An efficient code for a particular stimulus distribution will dedicate more resources to frequent stimuli and encode these with greater precision. The FI is a function of the stimulus and reflects the variation of the precision of the encoding across different values of the stimulus property.

A large body of work has studied how tuning parameters, such as tuning width and firing rate, are related to the FI~\cite{Zhang1999,Pouget1999,Harper2004,Durant2007,Ecker2011,Yarrow2012,Lin2015,Arandia2016}. The FI is proportional to the squared slope of the tuning curve when the noise is Gaussian and additive (Box 1). When the noise follows a Poisson distribution, the FI is  the squared slope of the tuning curve divided by the firing rate. Beyond single neurons and particular noise distributions, the FI provides a general stimulus-specific measure that characterizes coding fidelity of a neuronal population under various noise distributions. The FI sums across neurons, to yield the neural-population FI, if the neurons have independent noise. However, when the noise is correlated between neurons, the Fisher information is more difficult to compute~\cite{Zohary1994,Shadlen1996,Bair2001,Kohn2005,Cohen2009,Cohen2011}. This is still an active area of research~\cite{Abbott1999,Yoon1999effect,Nirenberg2003,Latham2005,Pola2003,Ecker2011,Moreno2014,Kafashan2021}, and some progress has been reviewed in ~\cite{Averbeck2006,Kohn2016}.

\subsection{Tuning determines geometry}
A tuning curve describes the encoding of a stimulus property in an individual neuron. When we consider a population of neurons, we can think of the response pattern across the $n$ neurons as a point in an $n$-dimensional space. Let's assume the tuning curves are continuous functions of the stimulus property. As we sweep through the values of the stimulus property, the response pattern moves along a 1-dimensional manifold in the neural response space. Now imagine that stimuli vary along $d$ property dimensions that are all encoded in the neural responses with continuous tuning curves. As we move in the $d$-dimensional stimulus-property space, the response pattern moves along a $d$-dimensional neural response manifold (Box 2).

\noindent\fbox{
\footnotesize{
    \parbox{0.98\textwidth}{
        {\bf BOX 2: Neural response manifold} \\
    A neural response manifold is a set of response patterns elicited by a range of stimuli~\cite{Seung2000,Dicarlo2007,Shenoy2013,Jazayeri2017}. In neuroscience,  we often think of a curved hypersurface of the neural response space within which each stimulus has its unique location. The mathematical notion of manifold implies that the set of neural response patterns is locally homeomorphic to a Euclidean space. This notion is particularly helpful when (1) the tuning curves are continuous, (2) the mapping between $d$-vectors of stimulus properties and $n$-vectors of neural responses is bijective, and (3) we neglect the noise. Under these conditions, the set of neural response patterns is homeomorphic to the particular Euclidean space spanned by the stimulus properties, and we can imagine mapping the stimulus space onto a hypersurface (which might be nonlinear, but will not intersect itself) in the multivariate response space.

    Consider a pair of neurons encoding orientation with sinusoidal tuning curves. Because both tuning curves are continuous, the response will vary continuously and cyclicly as the stimulus rotates. When the phases of the tuning curves are offset by $90^\circ$, the response manifold is a perfect circle (\textbf{a}). As the phase offset increases, the manifold gradually morphs into an ellipse with increasing aspect ratio (\textbf{b}, \textbf{c}). The shape of the manifold depends on the exact shape of the tuning curves. Panel (\textbf{d}) shows a crescent-like response manifold, induced by two von Mises tuning curves. For certain tuning functions, the response hypersurface may self-intersect (\textbf{e}). In the neighborhood of the self-intersection, the hypersurface is not homeomorphic to any Euclidean space, and thus not a manifold. However, adding one additional neuron (neuron 3 in \textbf{f}), disentangles the hypersurface, and renders it a manifold again. The width of bell-shaped tuning curves (\textbf{g} and \textbf{h}) substantially modulates the geometry of the manifold, with narrower tuning curves (\textbf{h}) leading to a manifold that is highly curved and requires a higher-dimensional linear subspace to explain most the variance (Fig.~\ref{fig:setup}, compare rows a and b, rightmost column).

    These examples neglect the presence of noise. When the responses are noisy, they may still form a manifold, but not a $d$-dimensional manifold, whose dimensions correspond to the $d$ stimulus properties of interest. Despite these caveats, the notion of manifold appears useful, because it enables us to see the response patterns elicited by a $d$-dimensional space of stimuli as a $d$-dimensional sheet in the multivariate response space~\cite{Seung2000,Dicarlo2007,Churchland2012,Shenoy2013,Jazayeri2017,Zhou2018,Ringach2019,Stringer2019}.

\centering
\includegraphics[keepaspectratio,width=1.01\linewidth]{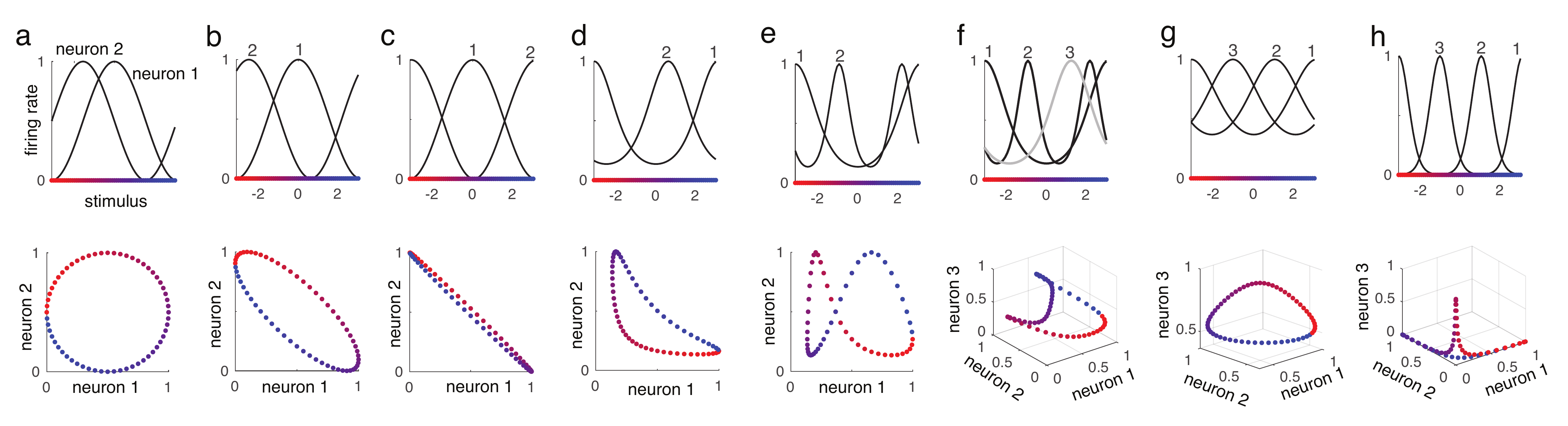}

}}}

The concept of a neural response manifold~\cite{Seung2000,Dicarlo2007} helps us relate a uni-dimensional or multidimensional stimulus space to a high-dimensional neural response space. To understand the code, we must understand the \textit{geometry} of the neural response manifold. As we saw above, the FI tells us the \textit{magnitude of the response pattern change associated with a local move in stimulus space} around any given location $\theta$. To define the representational geometry, however, we need a global characterization. We need to know the \textit{magnitudes of the response pattern changes associated with arbitrary moves} from any stimulus to any other stimulus. The geometry is determined by the distances in multivariate response space that separate the representations of any two stimuli. The tuning curves define the response patterns, from which we can compute the representational distances.

We could simply use the Euclidean distances $d(\theta_i,\theta_j)$ for all pairs $(i,j)$ of stimuli, which can be computed on the basis of the tuning functions alone to define the representational geometry:
\begin{equation}
    d(\theta_i,\theta_j) = \|\mathbf{f}(\theta_i)-\mathbf{f}(\theta_j)\|.
\end{equation}
However, we would like to define the representational geometry in a way that reflects the discriminability of the stimuli. The discriminability could be defined in various ways, for example as the accuracy of a Bayes-optimal decoder, as the accuracy or $d'$ of a linear decoder, or as the mutual information between stimulus and neural response. The Euclidean distance between stimulus representations is monotonically related to all these measures of the discriminability when the noise distribution is isotropic, Gaussian, and stimulus-independent. When the noise is anisotropic (and still Gaussian and stimulus-independent), we can replace the Euclidean distance with the Mahalanobis distance
\begin{equation}
    d_{Mahal}(\theta_i,\theta_j) = \sqrt{[\mathbf{f}(\theta_i)-\mathbf{f}(\theta_j)]^T\mathbf{\Sigma}^{-1}[\mathbf{f}(\theta_i)-\mathbf{f}(\theta_j)]} = \|\mathbf{\Sigma}^{-\frac{1}{2}}[\mathbf{f}(\theta_i)-\mathbf{f}(\theta_j)]\|,
\end{equation}
where $\mathbf{\Sigma}$ is the noise covariance matrix. The Mahalanobis distance will be monotonically related to the discriminability~\cite{Kriegeskorte2019}.

When the noise is Poisson and independent between neurons, a square-root transform can be used to stabilize the variance and render the joint noise density approximately Gaussian and isotropic (Box 1,~\cite{Bartlett1947}). How to best deal with non-Gaussian (e.g. Poisson) stimulus-dependent noise is an unresolved issue.

We have seen that the geometry of the representation depends on the tuning and the noise and can be characterized by a representational distance matrix. Changing the tuning will usually change the representational geometry. Consider the encoding of stimulus orientation by a population of neurons that have von Mises tuning with constant width of the tuning curves and a uniform distribution of preferred orientations. Since orientations form a cyclic set and the tuning curves are continuous, the set of representational patterns must form a cycle (Fig. 1a). However, two uniform neural population codes with different tuning widths will have distinct representational geometries.

With narrow tuning width (Fig. 1a), a given stimulus drives a sparse set of neurons. For intuition on this narrow tuning regime, imagine a labeled-line code, where each neuron responds to a particular stimulus from a finite set (one neuron for each stimulus). As we move through the stimulus set, we drive each neuron in turn. The set of response patterns, thus, is embedded in a linear subspace of the same dimensionality as the number of neurons. Note also that narrow tuning yields a ``splitter" code, i.e., a code that is equally sensitive to small and large orientation changes. Orientations differing by 30 degrees ($\pi/6$ radians) might elicit non-overlapping response patterns and would, thus, be as distant in the representational space as orientations differing by 90 degrees ($\pi/2$ radians, Fig. 1a, second-to-last column).

With wide tuning width (Fig. 1b), the geometry approximates a circle. It would be exactly a circle for two neurons with sinusoidal tuning with a $\pi/4$ shift in preferred orientation (where $\pi$ is the cycle period for stimulus orientation), and in the limit of an infinite number of neurons with sinusoidal tuning (with preferred orientations drawn from a uniform distribution). For von Mises tuning curves and finite numbers of neurons, wide tuning entails a geometry approximating a circle, with most of the variance of the set of response patterns explained by the first two principal components. Wide tuning yields a ``lumper" code, i.e., a code that is more sensitive to global orientation changes, than to local changes. Orientations differing by 30 degrees will elicit overlapping response patterns and would be much closer in the representational space than orientations differing by 90 degrees (Fig. 1b, right column).

If we make the gain or the width of the tuning curves non-uniform across orientations, the geometry is distorted with stimuli spreading out in representational space in regions of high Fisher information and huddling together in regions of low Fisher information (as illustrated in the RDM and MDS in Fig. 1c).

This section explained how a set of tuning functions induces a  representational geometry. We defined the geometry by its sufficient statistic, the representational distance matrix. The distances are measured after a transformation that renders the noise isotropic. We will see below that the distance matrix then captures all information contained in the code. Importantly, it also captures important aspects of the format in which the information is encoded. For example, knowing the geometry and the noise distribution enables us to predict what information can be read out from the code with a linear decoder (or in fact any given nonlinear decoder capable of a linear transform).

\subsection{Geometry does not determine tuning}
We have seen that the set of tuning curves determines the geometry (given the noise distribution). The converse does not hold: The geometry does not determine the set of tuning curves. To understand this, let us assume that the noise is isotropic (or has been rendered isotropic by a whitening transform). It is easy to see that rotating the ensemble of points corresponding to the representational patterns will leave the geometry unchanged. Such a rotation could simply exchange tuning curves between neurons, while leaving the set of tuning curves the same. In general, however, a rotation of the ensemble of points will substantially and qualitatively change the set of tuning curves (Fig. 1, compare b and d). The new tuning curves will be linear combinations of the original ones and, jointly, will induce the same geometry. 

If different sets of tuning curves can give rise to the same representational geometry, what information about the tuning is lost when considering the geometry? For example, the geometry doesn't tell us how the information about a given stimulus distinction is distributed across the neurons. The information could be concentrated in a small set of selective neurons or distributed across many neurons. However, the geometry is sufficient for determining how well any given distinction among the stimuli can be decoded (under the assumption of isotropic, stimulus-independent noise). 

In the previous section, we described the geometry as generated by the set of tuning functions. In this section, we saw that we can, conversely, think of a neuron's tuning curve as a projection of the ensemble of representational patterns onto a particular axis in the multivariate response space. Each neuron's tuning can be obtained by projecting the geometry onto a different axis. There are many different sets of axes (neurons) that can span the space and generate the same representational geometry.

\begin{figure}
\centering
\includegraphics[keepaspectratio,width=0.9\linewidth]{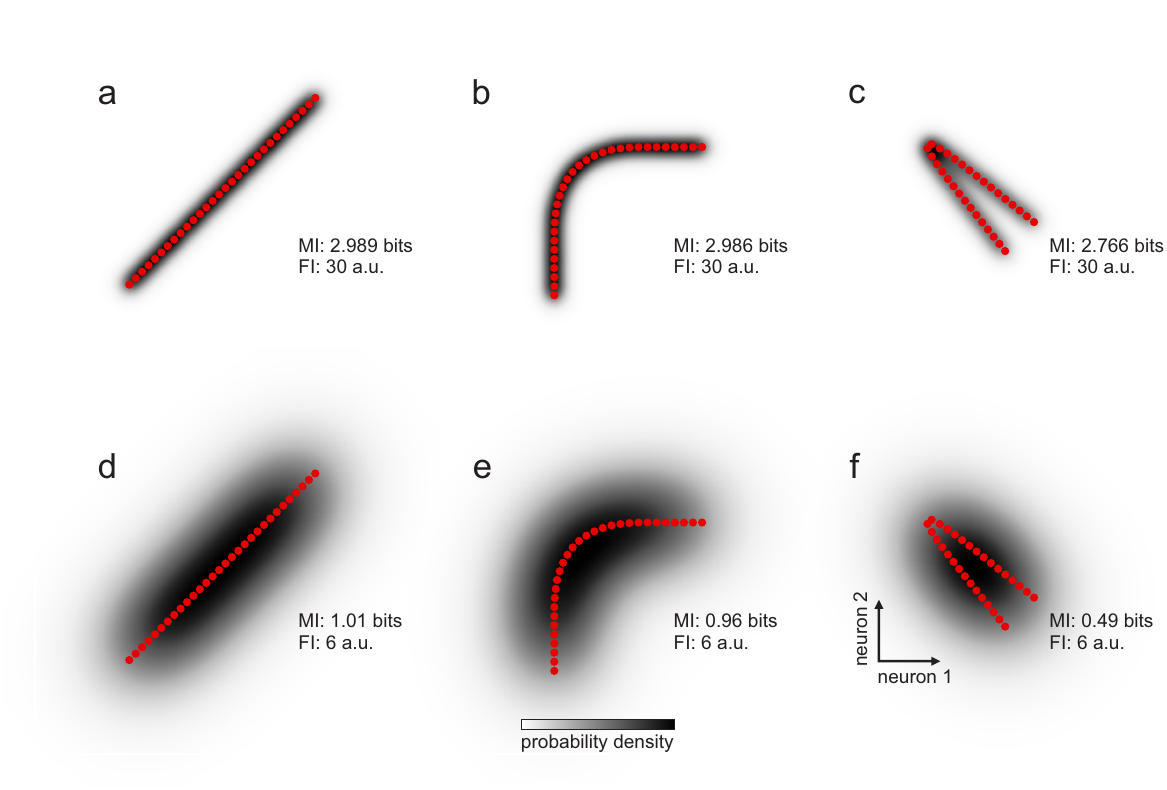}
\small
\caption{\textbf{The relationship between manifold geometry, Fisher information, and mutual information.} This toy simulation illustrates the encoding of a one-dimensional stimulus space by six different neural codes (a-f). The neural population comprises only two neurons, whose activities define the horizontal and vertical dimensions of each panel. For each of 30 equally spaced points in stimulus space, the mean response is plotted as a red dot. The neural response manifold is a continuous line (not shown) that passes through the red dots. The response distribution is shown in the background (grayscale) for a uniform prior over the stimulus space. The neural noise (Gaussian and isotropic here) is low in the top row (a-c) and high in the bottom row (d-f). The geometry of the manifold and the noise determine the Fisher information (FI) for each stimulus, the total FI, and the mutual information (MI) between stimulus and response. The total FI is proportional the length of the manifold divided by the standard deviation of the noise. The length of the manifold is the same in all codes (a-f). The total FI is 30 in the low-noise codes (a-c) and 6 in the high-noise codes (d-f). Within each noise level (top row, bottom row), the FI is the same, whereas the geometry and the MI change. This reflects the fact that the FI measures discriminability along the manifold, whereas the geometry and the mutual information measure discriminability among all stimuli. Folding the manifold (c, f) renders stimuli on opposite banks somewhat confusable, which lowers MI, but does not affect FI. The FI, thus, does not determine the geometry or the mutual information. Numerical values (lower right in each panel) show the MI and the total FI.}
\label{fig:FIandMI}
\end{figure}

\subsection{Geometry determines Fisher information and mutual information}

Fisher information and mutual information are usually computed based on tuning and noise properties. However the representational geometry provides powerful constraints on both FI and MI. In fact, if the noise is isotropic (or has been rendered isotropic by whitening), the geometry alone determines FI and MI, and we do not require the tuning curves to compute them.

\subsubsection{Fisher information} 
Consider $n$ neurons with tuning functions $f_i(\theta)$ (expressed as $\bf{f}    (\theta)$ for the population) and independent Gaussian noise. We assume the responses are scaled such that the noise has unit variance, without loss of generality. The Fisher information can be expressed as:
\begin{equation}
J(\theta)=\sum_{i=1}^{n} f_i'(\theta)^2= \norm{\frac{\partial\mathbf{f(\theta)}}{\partial \theta}}_2^2
\end{equation}

Note that the FI is the squared norm of the gradient of the response pattern with respect to the stimulus variable $\theta$. The FI, thus, reflects how far we move through representational space per unit of $\theta$. If we sweep through the values of $\theta$ at a constant rate, the FI is proportional to the speed of the corresponding trajectory along the neural manifold. The representational speed can change with the value of $\theta$, and so the FI is a function of $\theta$. The representational geometry defines the representational distances between pairs of stimuli. It therefore defines, in particular, the distance traveled as we move from stimulus value $\theta_i$ to the next stimulus value $\theta_{i+1}$. The representational geometry, thus, fully determines the FI.

So far we have been focusing on the FI as a function of the stimulus $\theta$. The overall FI can be defined \cite{Wei2015,Wei2016} as
\begin{equation}
J_{tot}=\int \sqrt{J(\theta)}d\theta~.
\label{eq:Jtot}\end{equation}
This quantity corresponds to the total length of the neural manifold $\mathbf{f}(\theta)$ which is a 1D curve in the neural response space (Fig. \ref{fig:FIandMI}). Because the overall FI is the length of the neural response manifold, it is invariant to re-parameterization of the stimulus. If the noise were heteroscedastic, this would need to be taken into account when calculating the FI.

\subsubsection{Mutual information}
The deterministic tuning functions $\mathbf{f}(\theta)$ together with a probabilistic noise model define the likelihood $p(\mathbf{r}|\theta) $, i.e. the probability of each possible response pattern $\mathbf{r}$ given the stimulus parameter $\theta$. We additionally need to specify a prior $p(\theta)$ to define the joint distribution $p(\mathbf{r},\theta) = p(\mathbf{r}|\theta) \cdot p(\theta)$. The distribution of the responses then obtains as  $p(\mathbf{r})=\sum_{\theta}p(\mathbf{r},\theta)$. The mutual information $I(\mathbf{r};\theta)$ between stimulus and response can be computed using the definition of the entropy and the conditional entropy (Box 1):
\begin{equation}
    I(\mathbf{r};\theta) = H(\mathbf{r}) - H(\mathbf{r}|\theta).
\end{equation}
The only transformations of the representation that preserve the geometry are rotations, translations, and reflections of the neural response manifold. These operations reparameterize $\mathbf{r}$, but  not in a way that changes the set of densities of the joint distribution. Instead the joint distribution is rotated and reflected with the geometry and the MI, a function of the joint density, is not altered.

Consider the special case of additive isotropic Gaussian noise. It is easy to show that $H(\mathbf{r}|\theta)=H(\epsilon)$, which is the entropy of the noise. Thus the MI is equal to the difference between the response entropy and the noise entropy. In other words, the MI is the portion of the response entropy that is not noise entropy: the stimulus-related portion of the response entropy. Both the response entropy $H(\mathbf{r})$ and noise entropy $H(\mathbf{r}|\theta)$ are invariant to rotations, translations, and reflections of the neural response manifold, thus all we need to compute the MI is the geometry and the noise.

The geometry of the neural representation can modulate the MI even when the FI is kept constant for all stimuli  (Fig. \ref{fig:FIandMI}, top row, bottom row). Here, we assume a uniform prior on stimulus $\theta$. We start with a straight-line neural response manifold (left). We then bend the line segment into a curved line (middle) and finally into a fold (right), while keeping the length of the manifold fixed and the density of the stimuli uniform along the manifold. The uniform distribution of the stimuli along the manifold means that stimuli remain equally discriminable from their neighbors in stimulus space. This is reflected in the FI, which remains unchanged for all stimuli across the three manifolds for a given noise level (row). However, the MI gets smaller as we fold over the manifold, because stimuli on opposite banks of the manifold come closer to each other in representational space and become confusable, especially at high noise (bottom row).

\subsection{Fisher information does not determine geometry}
We saw in Fig. \ref{fig:FIandMI} that the bending of the neural response manifold, a change of the global geometry, did not affect the FI. Each row in Fig. \ref{fig:FIandMI} shows three different neural codes, progressively bending the manifold, each of which has the same total FI and the same local FI for every stimulus, but a different representational geometry. The key insight here is that the FI, though it constrains the local geometry, does not constrain the global geometry of the manifold.

A complication in relating the FI to the geometry is that the FI is usually defined as a function of a continuous stimulus parameter, whereas the geometry is usually defined in terms of the distances among a finite set of stimulus representations. To align these two perspectives, we can consider a discrete approximation to the FI, based on the finite set of stimuli used to characterize the geometry. The red dots in Fig. \ref{fig:FIandMI}, provide such a finite stimulus set. Think of the dots as indexed from 1 to 30, with the linear interpolation defining the stimulus parameter for the FI. In the context of isotropic Gaussian noise, the FI is sum of squared slopes of the tuning curves, which is the squared representational distance between adjacent stimuli along the manifold. The FI as a function of stimulus index, thus, is defined by a \textit{subset} of the pairwise distances: those between two stimuli that are adjacent along the manifold. The FI, thus, only provides partial constraints on the representational geometry. For each stimulus, the FI only reflects the magnitude of the local within-manifold gradient of the stimulus parameter, whereas the geometry characterizes the distance to all other stimuli.

To deepen our intuition, let us return to the examples in Fig. \ref{fig:setup} a, b, where homogeneous neural populations encode orientation. In Fig. \ref{fig:setup} a and b, orientation is encoded by narrow (a) or wide (b) bell-shaped tuning functions. The FI is constant over orientations in both codes, but it is higher (assuming the same noise level) for narrow tuning. Doubling the width of the tuning curves doubles the number of neurons that respond to any given stimulus with the stimulus hitting either of the two sensitive flanks of the tuning curve. All else being equal, having more neurons change their activity with a small change of the stimulus would increase the FI. However, all else is not equal. Doubling the width also divides the slopes of the tuning curves by two everywhere, and the squared slopes by four. Overall, therefore, doubling the width of the tuning curves reduces the FI to half, and the total FI (i.e. the integral of the square root of the FI) by square root of two. For a one-dimensional stimulus space, therefore, narrowing bell-shaped tuning curves will increase the FI, as long as the population still covers the full range of stimulus values. Note that, for a two-dimensional variable like the location on a plane, doubling the width of bell-shaped tuning curves will quadruple the number of responsive neurons while still dividing each neuron's contribution to the FI by four. As a result, the FI will be invariant to tuning width for a two-dimensional variable~\cite{Zhang1999}.

The geometries induced by narrow and wide tuning are profoundly different (rightmost three columns: RDM, MDS, and representational distance as a function of orientation disparity). With narrow tuning, representational distance rapidly grows with orientation disparity (reflecting the high sensitivity to small changes and high FI). However, the representational distance also saturates quickly (with stimuli whose orientation differs by $\pi/2$ appearing no more distinct than stimuli differing by $\pi/4$). With wide tuning, representational distance saturates more slowly. With narrow tuning, the manifold occupies a high-dimensional linear subspace. With wide tuning, the manifold geometry approximates a circle, with almost all of the variance residing in a 2D linear subspace. These features of the geometry could be important to computation, but are not captured by the FI.

\subsection{Tuning determines perceptual sensitivity}

To understand neural computation, we must study not only how stimulus information is encoded in a region, but also how it is decoded by downstream neurons and reflected in behavior. If the population code we are studying formed the basis for perceptual decisions, we should be able to predict perceptual sensitivity that an ideal decoder could achieve on the basis of the tuning of the neurons and the noise. 

Predicting behavioral sensitivity requires a readout mechanism. We still do not know exactly how the brain reads out its codes to make perceptual decisions. Various decoders are theoretically possible. One fruitful and widely used approach is to assume an ideal observer (\cite{Green1966,Knill1996,Simoncelli2001,Geisler2011}), in other words, to assume that the decoder is optimal. This normative approach may not match what the brain does, but it enables us to predict the limits of behavioral performance given the code.

\subsubsection{General discrimination and perceptual estimation}
Bayesian ideal-observer models have been widely used to explain how perceptual estimates might be computed on the basis of the activity of a population of tuned neurons. The basic approach is to specify an encoding model (tuning functions and noise distribution) and simulate stochastic responses $\mathbf{r}\sim p(\mathbf{r}|\theta)$ from this model. One can then decode the stimulus from the simulated responses using Bayes rule, $p(\theta|\mathbf{r}) \propto p(\mathbf{r}|\theta)p(\theta)$, with the assumed prior distribution $p(\theta)$. One can simulate the ideal observer's perceptual estimates across many trials and characterize the estimation bias and variance to generate experimentally testable predictions~\cite{Tomassini2010,Girshick2011,Wei2015,Ruben2015}. This approach has been studied in computational work ~\cite{Sanger1996,Zhang1998,Oram1998,Ganguli2014} and has been used to understand characteristics of human and non-human primate perceptual behavior, including its prejudices (bias) and precision (variance)~\cite{Series2009,Gu2010,Graf2011,Bays2014,Wei2015}. Beyond predicting behavioral bias and variance, ideal-observer models have been used to explain the shape of the distribution of behavioral errors in detail, which is important, for example, to account for working-memory capacity~\cite{Bays2014,Zhang2008,Ma2014}. 

We may assume that the code is efficient ~\cite{Barlow1961} in that it maximizes stimulus-response MI under a set of resource constraints. A code's MI is the reduction of uncertainty about the stimulus that the ideal observer experiences upon seeing the neural response pattern (prior entropy minus posterior entropy). The resource constraints under which MI is maximized could be a limit on the number of neurons and a limit on number of spikes elicited by a stimulus in each neuron or in the entire population. Limiting neurons and spikes makes sense because each neuron must be developed, accommodated, and maintained in the brain, each spike comes at a cost in terms of energy~\cite{Levy1996,Laughlin2001,Balasubramanian2002}, and neurons cannot fire at arbitrarily high rates.

Finding a population of tuning functions that maximizes MI is a difficult optimization problem. Consider the case of $n$ neurons and an upper limit on the spike rate for each neuron. Optimizing MI will spread out the neural response patterns so as to fill the $n$-dimensional response hypercube. We may need to specify further that we want the tuning curves to be smooth or that we prefer a code that renders the ideal observer's posterior concentrated, so as to ensure that the code also performs reasonably in terms of the ideal observer's expected squared estimation error. The optimal code will then map the stimulus to a neural manifold. For a one-dimensional stimulus space, the manifold would snake through the $n$-dimensional hypercube forming a space-filling curve \cite{Ringach2010}. The optimization will trade off the length of the manifold against how close the manifold comes to itself as it bends and fills the response space. The length of the manifold is the total FI, which determines the confusability of similar stimuli. How close the manifold comes to itself determines the confusability of dissimilar stimuli. Both of these determine the ideal observer's ability to estimate and discriminate stimuli, and both are reflected in the MI. The tuning functions that would emerge from the maximization of MI (and the induced representational geometry) have yet to be fully investigated.

\subsubsection{Fine discrimination} 
Fine discrimination tasks enable us to establish thresholds and characterize how frequently similar stimuli are confused. However, they do not reveal how frequently dissimilar stimuli are confused. The performance in fine discrimination tasks is captured by the FI (Box 1). As we have seen above, assuming we know the noise distribution, the FI of each stimulus is determined by the tuning of the neurons. Theoretically, the FI provides a lower bound on the discrimination threshold for fine discrimination--a result that holds for both biased~\cite{Series2009} and unbiased~\cite{Seung1993} decoders. When we consider the FI and discrimination thresholds, we assume that only local distinctions between stimuli matter. This holds approximately when the neural manifold is smooth at the scale of the noise and never arcs back in on itself to render very different stimuli confusable (the scenario illustrated in the top row in Fig. \ref{fig:FIandMI}).

If we assume a minimal clearance that the neural manifold must maintain to itself as it fills the response space, there is a limit to the length it can achieve. If we are willing to assume that the confusability of dissimilar stimuli is negligible and that we can express the resource constraints in terms of a limit on the length of the manifold, then we can characterize the code in terms of its FI. The length of the neural manifold is proportional to the total FI $J_{tot}$ (Eq. \ref{eq:Jtot}) when the noise is isotropic and stimulus-independent. An efficient code will utilize the entire length of the manifold. The optimal mapping will depend on the frequency of the stimuli in natural experience, ensuring that the square root of the Fisher information is proportional to the stimulus prior~\cite{Mcdonnell2008,Ganguli2014,Wei2017,Clarke1994}. Stimuli that are more frequent in natural experience, thus, are granted more territory on the neural manifold, resulting in a uniform distribution of the stimuli on the response manifold. As a consequence, there is a direct relationship between the discrimination threshold and the stimulus prior~\cite{Ganguli2014,Wei2017}: $\delta (\theta) \propto \frac{1}{p(\theta)}$. This relationship quantitatively captures the intuition that stimuli that appear more frequently should be better encoded, resulting in greater perceptual sensitivity. The quantitative predictions might explain, for example, the well-established oblique effect in orientation discrimination~\cite{Appelle1972}. 

\begin{figure}
\centering
\vspace{-4mm}
\includegraphics[keepaspectratio,width=0.35\linewidth]{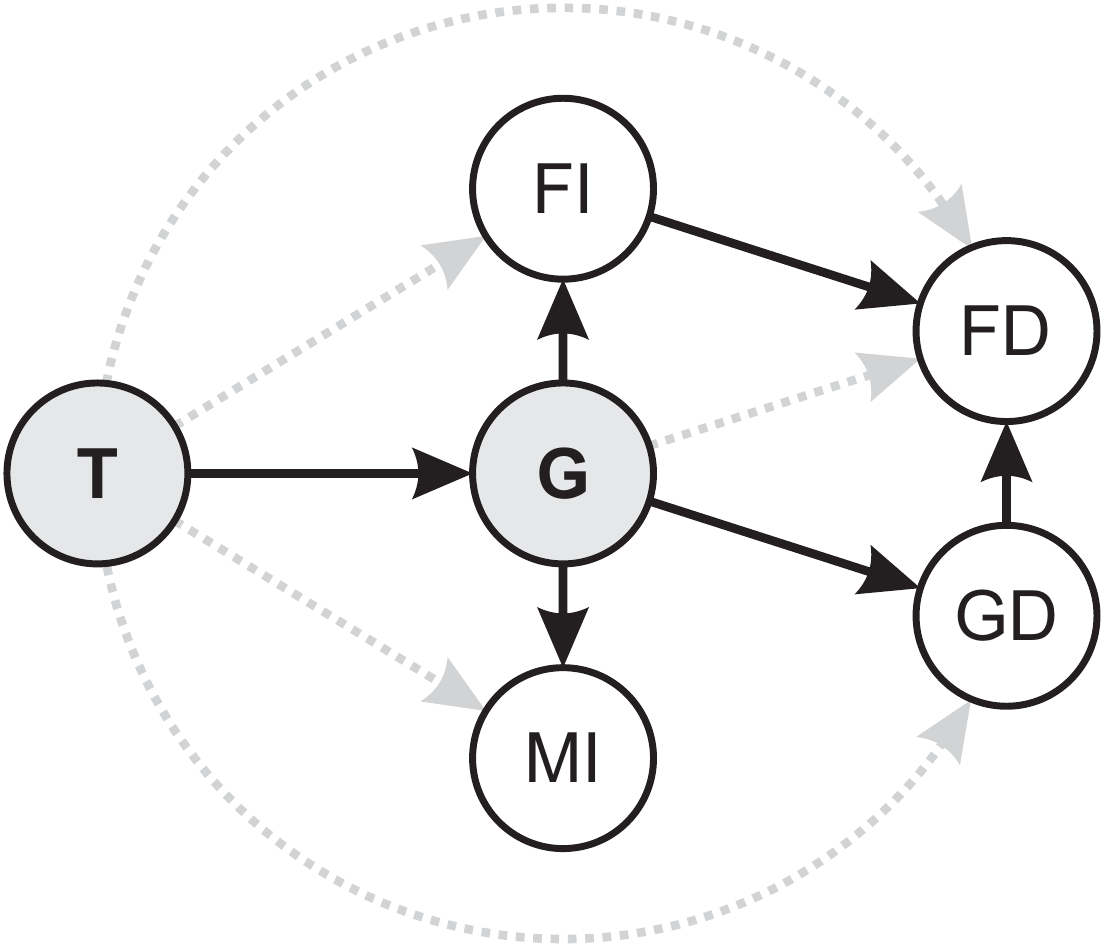}
\small
\caption{\textbf{The relationships among tuning, geometry, Fisher information, mutual information, fine discrimination, and general discrimination.} This graphical model shows the dependencies among tuning (T), geometry (G), Fisher information (FI), mutual information (MI), general discrimination (GD), and fine discrimination (FD). A directed edge from A to B means that A is sufficient to determine B, assuming always that we are given a model of the noise. The absence of a directed edge from B to A means that B is not sufficient to determine A. Note that the total graph (including all directed edges, solid black and dotted gray) is acyclic, indicating that there are no equivalence relationships in the graph. The skeleton of solid black directed edges captures the key relationships. The dotted gray directed edges are implied in the solid black directed edges by transitivity.}
\label{fig:graph}
\end{figure}

\subsection{Geometry determines perceptual sensitivity}
We can estimate the ideal observer's perceptual sensitivity without knowing the tuning curves. Knowing the representational geometry is sufficient to predict the ideal observer's performance in binary discrimination and perceptual estimation tasks. As before, we assume that the distances used to define the geometry have been measured after a transform of the response space that renders the noise isotropic and stimulus-independent.

\subsubsection{Binary discrimination} 
The performance of an ideal observer in a binary discrimination task is determined by the representational distance between the two stimuli. This can be construed as a generalization of signal detection theory~\cite{Green1966} from a univariate to a multivariate response. When the two points are close in the representational space, \textit{general discrimination} reduces to the special case of \textit{fine discrimination}. We have seen above that the geometry determines the FI and that the FI, in turn, determines the ideal observer's performance at fine discrimination. However, the geometry determines the discriminability of arbitrary pairs of stimuli.

Beyond pairs of stimuli, the geometry determines how well any dichotomy of the stimuli can be read out by an ideal observer or, more generally, by any given more restricted linear or nonlinear decoder that has access to the entire population and is capable of affine transformations \cite{Kriegeskorte2019}. This holds when the representational distances have been measured after transforming the space to render the noise isotropic and stimulus-independent.

\subsubsection{Perceptual estimation}
We saw above how a Bayesian ideal observer might estimate the stimulus parameter $\theta$ from a response pattern across a population of tuned neurons. To compute the ideal observer's estimation performance, we needed to know the tuning functions and the noise distribution, which jointly determine $p(\mathbf{r}|\theta)$. Once the response space has been transformed to render the noise isotropic and stimulus-independent, however, the geometry is sufficient to compute the ideal observer's performance at estimation.

Any two sets of tuning functions that induce the same geometry are related by a sequence of translations, rotations, and reflections of the ensemble of representational patterns. If the code is altered by a sequence of such transformations that maps $\mathbf{r}\rightarrow \mathbf{r}'$, the ideal observer needs to update the readout model (undoing the transformations by an inverse linear transform), but will then obtain the same posterior $p'(\theta|\mathbf{r}') = p(\theta|\mathbf{r})$ for any given response pattern. This point is elaborated in Supplementary Section \ref{sec:rotinv}. The geometry, thus, determines the ideal observer's error distribution. How the error distribution depends on the geometry and the stimulus can be simulated. However, a closed-form expression of this relationship is not currently available.

One may wonder whether knowing the FI suffices to predict the distribution of estimation error. The answer, in general, is no. The FI provides a lower bound on the variance of the error distribution. However, the error distribution may not be Gaussian, and FI by itself is insufficient to determine the shape of the error distribution. 

\section{Reconsidering notable neural codes from the dual perspective}

\begin{figure}
\centering
\vspace{-5mm}
\includegraphics[keepaspectratio,width=0.9\linewidth]{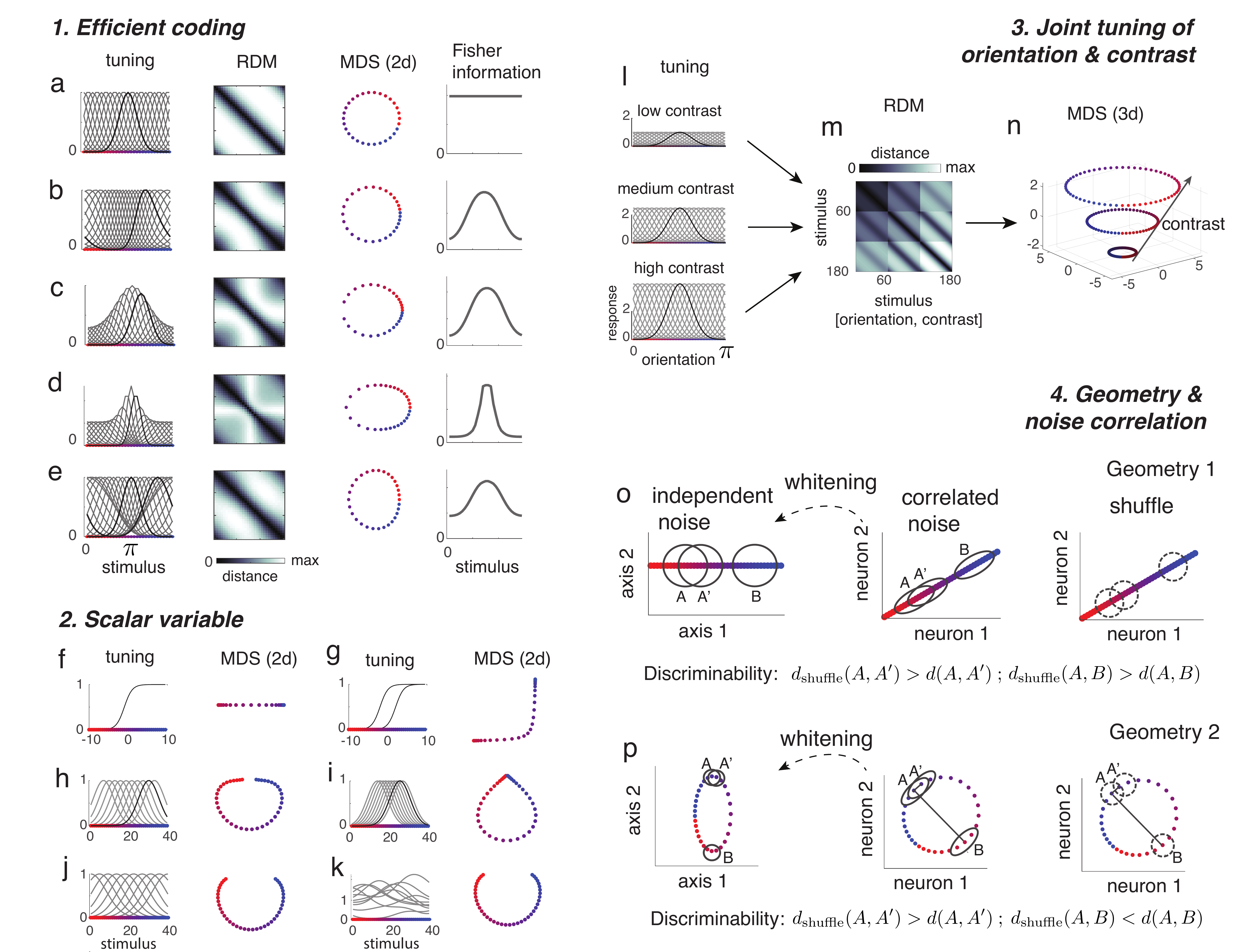}
\vspace{-4mm}
\small
\caption{\textbf{Example applications}. (a-e) Efficient coding. (\textbf{a}) A coding configuration with homogeneous, von Mises tuning curves. Panels (b-e) illustrate four efficient coding schemes for a stimulus prior that is smooth and peaks at $\pi$ (most frequent stimuli shown as purple dots). (\textbf{b}) Nonlinear warping of the stimulus space yields asymmetric tuning curves, conserving the manifold while redistributing the stimuli within it, so as to optimally allocate the FI according to the prior~\cite{Ganguli2010}. (\textbf{c}) Increasing the gain of neurons preferring stimuli close to the peak of the prior, without altering the shape of their tuning, achieves an identical reallocation of the FI, but manifold geometry and MI differ from (b). (\textbf{d}) The gain is scaled up for neurons that prefer stimuli closer to the peak of the prior ($\pi$). In addition, their tuning is sharpened, so as to conserve the area under each tuning curve~\cite{Li2018}. (\textbf{e}) Sharpening of the tuning of neurons preferring stimuli with higher prior density (without gain adjustment) can achieve a similar reallocation of the FI~\cite{Girshick2011}. (f-k) Alternative codes for a scalar, non-cyclic stimulus variable. (\textbf{f}) A single-neuron code with monotonic tuning curve and the corresponding neural manifold. (\textbf{g}) Similar to (f) but for a pair of neurons with shifted tuning curves. (\textbf{h}) Encoding a scalar with a population of bell-shaped tuning curves. (\textbf{i}) Like (h), but tuning curves don't tile out the extremes, so extreme stimuli elicit no response. (\textbf{j}) Tuning curves with preferred stimuli uniformly and completely covering the entire stimulus domain. (\textbf{k}) A rotated and translated version of the code in (j). The codes in (j) and (k) have different tuning curves, but identical geometry, FI, and MI. The representational distance grows monotonically with stimulus disparity in codes (f) and (g), but not in codes (h-k). (l-n) A neural population that jointly encodes stimulus orientation and contrast.  (\textbf{l}) Each neuron exhibits joint tuning for orientation and contrast. (\textbf{m}) The representational dissimilarity matrix (RDM) for a set of 180 stimuli (60 orientations $\times$ 3 contrast levels). (\textbf{n}) MDS embedding visualizes the geometry of the joint coding scheme. (o, p) Geometry and noise correlation. When the two stimulus representations differ along high-variance dimensions of the noise, the stimuli are less discriminable than when they differ along low-variance dimensions of the noise. (\textbf{o}) For a linear manifold, the effect of noise correlation on local discriminability (example $d(A,A')$) is the same across the stimulus space. The same goes for global discriminability (example $d(A,B)$). (\textbf{p}) For a curved manifold, the effect of noise correlation on discriminability (both local and global) depends on the stimulus values because the signal orientation varies relative to the noise orientation. Artificially shuffling trials independently for each neuron increases decoding accuracy when the representational difference vector is oriented along a high-variance dimension of the noise (panel (o) for $d(A,A')$ and $d(A,B)$, panel (p) for $d(A,A')$), and decreases decoding accuracy when the representational difference vector is oriented along a low-variance dimension of the noise (panel (p) for $d(A,B)$).}
\label{fig:app}
\end{figure}

We now illustrate how the proposed framework (Fig.\ref{fig:graph}) illuminates efficient coding, the coding of one-dimensional, two-dimensional, and higher-dimensional variables, and the effect of noise correlations. We use theoretical examples that are relevant for understanding the encoding of perceptual and motor variables, such as contrast, orientation, spatial frequency, speed, and direction. An additional higher-cognitive example of particular interest is the encoding of spatial environments by grid cells, which we address in Supplementary Section 4.2. In all these cases, considering the dual perspective of tuning and geometry gives us a deeper understanding and interesting insights.

\subsection{Efficient coding}
A prominent and long-standing hypothesis in neuroscience is that a neural system should adapt to the statistical structure of the sensory data to achieve an efficient code ~\cite{Barlow1961,Attneave1954}. A code is considered efficient if it provides much information about the stimulus (high MI) under constraints on the number of spikes and neurons. The efficient coding hypothesis provides a rationale that may explain why neurons employ particular tuning functions to represent a variety of sensory and non-sensory variables~\cite{Barlow1961,laughlin1981,Atick1992,Olshausen1996,Bell1997,Hateren1998,Simoncelli2001,Schwartz2001,Lewicki2002,Wei2015grid,Mosheiff2017}. Efficient coding has also helped explain human behavior in psychophysical tasks~\cite{Girshick2011,Ganguli2014,Wei2015,Wei2017,Li2018,Polania2019}.

The efficient-coding hypothesis enables us to make predictions about the shape of the tuning functions. However, it is difficult to jointly optimize all tuning functions for a population of neurons to maximize MI or FI. Past theoretical studies have therefore introduced simplifying assumptions, including the assumption of unimodal tuning and particular functional forms for the tuning curves. Consider the challenge of  encoding a cyclic stimulus variable, such as orientation, when the prior is uniform. In this context, it is often further assumed that the tuning functions are homogeneous: Each unimodal tuning function has the same shape and the neurons' preferred stimuli are uniformly distributed. An interesting question is how such a code should be adjusted when the prior is non-uniform.

Intuitively, an efficient code should allocate more resources to more frequent stimuli (e.g. more neurons, spikes, or volume within the multivariate response space or neural response manifold). An optimal allocation can be achieved by nonlinear warping of the stimulus parameterization, so as to render the prior uniform for the reparameterized stimuli. The optimal tuning functions will then be homogeneous for the reparameterized stimuli, and warped for the original  parameterization~\cite{Ganguli2014}. The warping renders the shape of the tuning asymmetric and the density of tuning peaks nonuniform, but the optimal code has constant gain (i.e., all neurons have the same peak firing rate)~\cite{Ganguli2014}. Instead of warping (nonlinear stimulus reparameterization), we can vary the width of symmetric (e.g., von Mises) tuning functions across neurons~\cite{Girshick2011} or the gain ~\cite{Wei2015}, or both~\cite{Li2018}.

Considering how these different adjustments of the tuning functions affect the representational geometry reveals their effect on FI, MI, and perceptual discriminability. For each coding scheme, Fig.~\ref{fig:app} shows the tuning curves, the geometry (RDM, MDS), and the FI profile. Interestingly, compared to the homogeneous population, warping the tuning curves by a nonlinear stimulus reparameterization does not change the neural manifold. Instead, warping moves the points along the manifold, so as to allocate the FI according to the prior. Varying the tuning width and gain both change the shape of the neural manifold relative to a neural population with homogeneous tuning. The adaptive-gain model (Fig.~\ref{fig:app}d, ~\cite{Li2018}) can lead to a prominent expansion of the geometry near the peak of the prior, where the sharpening of the tuning curves and the increase of the gain coincide (preserving the area under each curve). Even subtle sharpening and gain increase are very potent in combination, and can enhance FI and MI substantially. 

The four schemes of heterogeneous tuning (warping, varying gain, varying width, varying both) all increase the FI around the peak of the prior (Fig.~\ref{fig:app}b-e, purple dots more widely spaced in MDS). Note that the warped code and the code with varying gain have identical FI profiles here. However, the shape of their neural manifolds is distinct, as reflected in the representational geometry. The geometry reveals different predictions that the codes make about the perceptual discriminability under high noise. For example, the most frequent orientation will be especially discriminable from the opposite orientation for the code with varying gain, but not for the warping code (purple dots in MDS in Fig.~\ref{fig:app}b, c). 

\subsection{Encoding a non-cyclic scalar variable}

The examples considered so far concerned encoding of cyclic variables. Of course, the brain is capable of representing non-cyclic scalar variables, such as contrast~\cite{laughlin1981,Anderson2000}, spatial frequency~\cite{Campbell1969,de1982}, and the speed of a moving object~\cite{Maunsell1983,Nover2005}. In principle, these variables could be represented either by a single neuron or multiple neurons with monotonic tuning curves (as in the contrast code of a fly H1 neuron~\cite{laughlin1981}) or by a population of neurons with bell-shaped tuning curves (as in the spatial-frequency code in V1~\cite{Campbell1969,de1982} and the speed code in MT~\cite{Maunsell1983,Nover2005}).

What are the implications of these various coding schemes for the representational geometry? First, if a variable is represented using a single neuron with monotonic tuning, the neural manifold is simply an interval on a straight line, bounded by the range of the firing rate (Fig.~\ref{fig:app}f). Varying the shape of the tuning curve will not change the manifold, rather it will redistribute the points by moving them along the manifold. With two neurons, each exhibiting a monotonic tuning curve, the representation will be a curved manifold in general (Fig.~\ref{fig:app}g). The curvature depends on the overlap of the stimulus domains in which the two neurons are responsive. When all tuning curves are monotonic, the representational distance grows monotonically as we move away from any stimulus on the manifold. Consider the scenario where the tuning curves are monotonic, smooth, and homogeneous (i.e., tuning curves are all of the same shape, e.g. sigmoidal, with a uniform distribution of thresholds). As we move away from a stimulus, the representational distance grows linearly at first, within the range matching the transition width of the tuning curves. However, as we move further in stimulus space and more and more neurons switch on (or off), the representational distance grows as the square root of the stimulus disparity (as the disparity is proportional to the number of additionally activated or deactivated neurons).

Now consider a population of bell-shaped tuning curves encoding a non-cyclic scalar variable. The resulting neural manifold strongly deviates from linearity (Fig.~\ref{fig:app}h-j). Small stimulus disparities (relative to the tuning width) will again be proportionally reflected in representational distances. However, two stimuli with large disparity will be represented by non-overlapping neural populations, so further increasing the stimulus disparity will not further increase the representational distance. If the tuning curves tile the entire domain homogeneously, the representational distance will never drop as we move away from a stimulus on the manifold. If responses peter out at the extremes of the stimulus domain, however, the manifold's extremes will converge in the representational space (Fig.~\ref{fig:app}h-j). This suggests an intriguing perceptual hypothesis: that extrema of a stimulus scale are perceptually similar~\cite{Stocker2009b}. Bell-shaped tuning and monotonic tuning, thus, induce fundamentally different representational geometries. Each geometry reveals how discriminable the code renders different pairs of stimuli, with monotonic tuning emphasizing the contrast between the extremes.

If we continually change the stimulus, sweeping through its scalar parameter space, the neural response pattern follows a trajectory that traces out the manifold. Equivalently, we may consider the case where the non-cyclic scalar variable encoded is time. Each tuning curve then effectively describes the time course of one neuron's firing rate. A set of neurons with bell-shaped tuning will be driven in sequence, waxing and waning in overlapping windows of time~\cite{Harvey2012}. Sequential activation of neurons with sufficient temporal overlap can yield a rotation-like trajectory (Fig.~\ref{fig:app}h-j, ~\cite{Lebedev2019}). 

A perfectly circular trajectory traversed at constant rate requires (1) that each neuron's activity oscillate sinusoidally at the frequency of revolution and (2) that the sum across neurons of each neuron's squared deviation from its mean activity be constant over time (and equal to the squared radius of the circle). These two properties are both necessary for a perfectly circular trajectory traversed at constant rate, and jointly sufficient. At a minimum, such exact rotational dynamics requires two neurons (with $90^{\circ}$ phase offset between their sinusoidal tuning curves). It can also be achieved with three or more neurons of varying amplitudes and phases. In particular, a large neural population with sinusoidal tuning of the same frequency will exhibit circular dynamics when the phases are uniformly distributed and the amplitudes constant.

Rotation-like dynamics has been reported in motor cortex of macaque monkeys when performing reaching tasks~\cite{Churchland2012,Michaels2016}. Neurons ramping up their activity at different latencies will create a curved representational trajectory. Rotation-like dynamics obtains when neurons also ramp down again, waxing and waning at similar frequency and peaking in sequence at a range of phases (Fig.~\ref{fig:app}k, which illustrates rotational dynamics identical to that in panel j ~\cite{Lebedev2019}).

\subsection{Joint tuning of multiple stimulus attributes}

Neurons can represent multiple stimulus attributes simultaneously. In primates, for example, many neurons in primary visual cortex (V1) are selective for stimulus orientation, contrast, and spatial frequency simultaneously~\cite{devalois1990,Anderson2000,Nauhaus2012}; neurons in area MT are jointly tuned to speed and direction of visual motion~\cite{Maunsell1983}. 

We illustrate the implications of joint tuning on the representational geometry using a neural population jointly tuned to orientation and contrast. We assume that individual neurons exhibit von Mises (i.e. circular Gaussian) tuning to orientation for every contrast level, and the gain (but not the preferred orientation) depends on the contrast. This simple model is consistent with previous experimental studies in macaque V1~\cite{Sclar1982,Anderson2000}. We simulated representations of stimuli of three contrast levels (low, medium, high) and 60 equally-spaced orientations, resulting in $60\times3=180$ representational states. We performed MDS to visualize the geometry (Fig.~\ref{fig:app}l-n).
For each contrast level, the neural manifold encoding the orientation approximates a circle (Fig.~\ref{fig:app}n). The roughly circular geometry results from the relatively wide tuning curves (cf. Fig.~\ref{fig:setup}a,b). Changing the contrast scales the orientation manifold uniformly about the origin of the response space. The joint manifold for orientation and contrast therefore has the geometry of a cone. Because high contrast yields a larger circle, the total FI for orientation grows with contrast under this multiplicative gain model.

Many other forms of joint tuning could be considered. One interesting case is the representation of composite stimuli, such as compound gratings, by a neural population.
Experimentally, it is well established that the components of a complex visual stimulus interact with each other. These interactions have been modeled by various forms of gain control~\cite{Morrone1982,Albrecht1991,Heeger1992,Deangelis1992a,Sillito1996,Carandini1998,Grill2006,Carandini2012,Benucci2013,Alink2018}, such as surround suppression and cross-orientation suppression. Recent work investigated how visual masking affects the representational geometry, leading to the novel insight that the effect of masking can be captured by an affine transformation of the geometry ~\cite{Ringach2010,Tring2018,Ringach2019}.

\subsection{Geometry and noise correlation}
Noise correlation (NC) affects the discriminability of stimuli in the neural code. This issue has received much attention in systems and computational neuroscience during the past two decades ~\cite{Zohary1994,Abbott1999,Averbeck2006,Averbeck2006b,Kohn2016,Nogueira2020}. The central insight is that the discriminability of two stimuli depends on the  overlap of their response distributions ($p(\mathbf{r}|\theta_1)$, $p(\mathbf{r}|\theta_2)$). When the noise is correlated (thus not isotropic), the discriminability of two stimuli depends not only on the strength of the signal (i.e. the norm of the difference vector between the two response pattern means), but also on the orientation of signal (i.e. the orientation of the difference vector) relative to the noise distribution~\cite{Averbeck2006,Kohn2016}. When we consider fine discrimination, the FI depends on the local orientation of the neural manifold relative to the multivariate distribution of the noise. For fine and coarse discrimination, when the signal is oriented along axes of high noise variance, NC will reduce the discriminability of stimuli (Fig.~\ref{fig:app}o, stimuli A and A'). When the signal is oriented along axes of low noise variance, NC will increase the discriminability of stimuli (Fig.~\ref{fig:app}p, stimuli A and B).

When we ask how NC affects stimulus discriminability, we are comparing the actual code to an imaginary alternative code in which the noise distribution for each neuron is the same, but the noise is not correlated between neurons. In data analysis, this alternative code without NC can be simulated by shuffling the trials independently for each of a set of simultaneously recorded neurons~\cite{Averbeck2006}. Shuffling will improve linear decoding accuracy when the signal and noise axes are aligned, and it will hurt decoding accuracy when signal and noise axes are misaligned. In the former case shuffling helps decoding by providing access to information from multiple trials. In the latter case shuffling hurts decoding by removing information about simultaneous activity of different neurons, which a linear decoder would have used to cancel the noise.

The geometrical perspective gives us a simpler understanding of this issue. Rather than considering an imaginary alternative code in which the response patterns are identical, but the noise has been artificially decorrelated, we linearly transform the response space (the signal and the noise \textit{together}), so as to whiten the noise~\cite{Hyvarinen2009}. Such a noise-whitening linear transform conserves linear decodability, FI, and MI, and renders the representational distances monotonically related to the discriminabilities of the stimuli (Fig.~\ref{fig:app}o, p). Consider a two-neuron code in which the original neural manifold is a circle, and there is positive NC between the two neurons (Fig.~\ref{fig:app}p). How does NC affect the discriminability of stimuli A and A'? Because the NC is aligned with the signal (response pattern A minus response pattern A'), the discriminability is smaller than for the shuffled code~\cite{Averbeck2006} (Fig~\ref{fig:app}p, right panel). By contrast, the discriminability of stimuli A and B is larger than for the shuffled code because the NC is orthogonal to the signal in this case. Considering the representational geometry (after whitening of the noise, Fig.~\ref{fig:app}o, p, left panels) clarifies these effects of the NC. Stimuli separated orthogonally to the axis of highest noise variance have greater representational distances, revealing their greater discriminability.

NC renders some directions in the response space more noisy than others. To maximize the FI, the neural manifold should flow along directions of low noise variance. A recent study found evidence consistent with this idea in mouse V1~\cite{Rumyantsev2020}. The worst-case scenario is that the manifold is aligned with the noise everywhere. The displacements caused by noise, then, look just like signal and confound the encoding as if the noise was already present in the input. In this scenario, the NC for a pair of neurons is proportional to the product of the derivatives of their tuning curves at the stimulus value. Since information already missing from the input cannot possibly be recovered from the code, such so-called \textit{differential} noise correlations limit the FI that the code can achieve, no matter how many neurons we allow~\cite{Moreno2014}. There is some evidence that differential noise correlations are present in mouse V1~\cite{Rumyantsev2020,Kafashan2021} and frontal cortex~\cite{Wimmer2014,Bartolo2020}.

The perspective of representational geometry suggests that we should consider not just the orientation of the manifold (the tangent), but also all other directions along which stimuli are separated (the cords) (Fig~\ref{fig:app}p). The local orientation of the manifold relative to the correlated noise affects the fine discriminability (as measured by the FI). The orientation of cords relative to the correlated noise affects the discriminability of stimuli represented in different parts of the manifold (and the overall MI). Neural manifolds tend to curve, so the effect of NC on local and coarse discrimination can be different and even opposite~\cite{Lin2015}. When the response space is high-dimensional~\cite{Stringer2019}, in particular, we may need to consider the full geometry of the neural manifold to understand the effect of NC.

\section{Current challenges}

Brain computation serves to extract and exploit behaviorally relevant information. The dual perspective of neural tuning and representational geometry provides a fuller picture of neural codes and how the encoded information and its format changes across spatial and temporal stages of processing. 

We considered the encoding of uni- and multidimensional variables from both perspectives. The tuning reveals how individual neurons encode information. The geometry abstracts from the particular set of tuning functions that encode the information. It summarizes the content and format of the code from the perspective of a downstream decoder that has access to the entire population and is capable of an affine transform~\cite{Kriegeskorte2019}. Whether the geometry is all that matters for computation is an important theoretical and empirical question. We have seen that two population codes with distinct sets of tuning functions can induce the same representational geometry, thus encoding the same information in a similar format. Why then might a brain develop one set of tuning functions rather than another? The particular tuning curves chosen to implement the representational geometry might matter for the following reasons:
\begin{itemize}
\item \textbf{Hypercubical range}: Each neuron has a limited range of firing rates from 0 to some maximum rate, yielding a hypercubical population response space. The optimal neural manifold geometry may fit into the hypercubical response space in some orientations, but not in others.
\item \textbf{Metabolic cost}: Neural systems may prefer codes with lower metabolic cost~\cite{Laughlin1998}. This would favor tuning functions with lower expected total number of spikes~\cite{Olshausen1996} (\textit{e.g.},  Fig.~\ref{fig:setup}b over Fig.~\ref{fig:setup}d), which nestle the manifold into the corner at the origin of the hypercubical response space.
\item \textbf{Encoding computations}: Neural tuning results from feedforward and recurrent computations. The encoding may be easier to compute or to learn for certain tuning functions. For instance, optimizing a recurrent neural network to track the animal's head orientation by angular integration has been shown to yield bell-shaped head-direction tuning~\cite{cueva2020}.
\item \textbf{Decoding computations}: Downstream regions reading the code to generate a percept or motor response may be better served by certain sets of tuning functions. For example, if readout neurons can only read a portion of the population, they will benefit from concentration of the information they need in that portion of the code \cite{Kriegeskorte2019}.
\end{itemize}

An exciting approach for future studies is to postulate a functional objective and a set of biological constraints, and to predict, by optimization, the representational geometry and/or its implementation in a set of tuning functions. This approach can enable us to address \textit{why} the brain employs certain geometries and \textit{why} it implements these geometries in particular tuning functions.

An important technical and theoretical issue is how to best quantify the representational geometry. If we transform the responses such that the noise is isotropic, Gaussian, and stimulus-independent, the Euclidean distance will reflect the discriminability for each pair of stimuli. Current studies use linear transforms and variance stabilization to render the noise approximately isotropic (Box 1). However, these transformations do not solve the problem in general. It is presently unclear whether it is always possible to make the noise isotropic by transforming the responses and how the transformation should be defined, in particular when the neurons have complex structured noise.

If no transform can be found that renders the noise isotropic, Gaussian, and stimulus-independent, one way forward is to give up on Euclidean geometry. We may simply define the dissimilarity for each pair of stimuli as the discriminability of the two stimuli, taking into account the particular noise distributions, which may differ between stimuli~\cite{Amari2000}. It is an interesting question how discriminability should be defined in this context. One could define it as the divergence between the two response distributions associated with a pair of stimuli (e.g., the Bhattacharyya distance~\cite{Bhattacharyya1943} or the symmetrized Kullback-Leibler divergence~\cite{Johnson2001}), or as the accuracy of the Bayes-optimal decoder or of some restricted class of readout mechanisms (e.g., linear decoders).

Manifold learning techniques may also help improve the characterization of the representational geometry under more relaxed assumptions about the noise~\cite{Tenenbaum2000,Roweis2000}. Recently, there has been a surge of interest in characterizing neural response manifolds without necessarily modeling the tuning of responses to properties of the stimulus or experimental condition~\cite{Chapin1999,Briggman2005,Macke2011,Mante2013,Shenoy2013,Sadtler2014,Pandarinath2018,Minxha2020}. 
Some of the methods proposed aim to characterize the topology, rather than the geometry of the mani\-fold. Unsupervised methods can recover the topology of the set of neural population activity patterns that occur during task performance, wakeful rest, or sleep. In order to characterize the topology, these methods exploit neighbor relationships in the neural response space (\textit{e.g.}, ~\cite{Singh2008,Giusti2015,Chaudhuri2019}) and in some cases also in time (\textit{e.g.},\cite{Low2018}) as the animal's brain traverses its state space.

A promising approach is to estimate the geometry of the high-dimensional neural responses by leveraging a low-dimensional latent space~\cite{Low2018,Zhou2020}. Such methods may enable us to recover the representational geometry in the latent space, such that the overlap of stimulus-independent isotropic Gaussian noise distributions in latent space reflects the overlap of the response distributions in the high-dimensional neural response space. The distances in latent space would then be monotonically related to the response-pattern discriminabilities.

The representational geometry likely reflects the statistical structure of the world and the behavioral implications of the represented content. Our examples assumed a uniform prior distribution over the stimulus variable. However, the brain's inferences are known to account for non-uniform prior distributions in perception and cognition~\cite{Knill1996,Langer2001,Weiss2002,Adams2004,Griffiths2006,Stocker2006,Tomassini2010,Girshick2011,Wei2015}. Decisions and control, similarly, require the weighing of risks and rewards~\cite{Kording2006}. How prior expectations and reward implications are reflected in representational geometries is an exciting direction for future research. A working hypothesis is that detailed representational geometries arise from a genetically determined blueprint through learning, and that they come to reflect the statistical regularities of the environment, which might be learned even without supervision or reinforcement~\cite{Gibson1979,Shepard1984,Shepard2001,Wiskott2002,Berkes2005}, and ultimately the behavioral demands and the reward structure of the environment~\cite{Shepard1984}.

Neural representational geometries may relate to \emph{conceptual spaces}~\cite{Gardenfors2004}, an abstraction used in cognitive science to capture the structure of mental models~\cite{Fechner1860,Thurstone1927,Attneave1950,Ekman1954,Beals1968,Shepard1984,Shepard1964b,Krantz1975,Shepard1987,Nosofsky1986}. To quantitatively reconstruct conceptual spaces, researchers have used multidimensional scaling  based on generalization behavior and similarity judgements~\cite{Ekman1954,Shepard1962,Torgerson1965,Tversky1977,Shepard1980,Tversky1982,Nosofsky1986,Shepard1987,Ashby1988,Kriegeskorte2012a,Hebart2020}. Neural and cognitive representational spaces should not be naively equated, but it is possible that they are closely related~\cite{Mur2013,Cichy2019}. Understanding their relationship in particular domains would help connect cognitive science to neuroscience.

Most studies of neural representation have focused on either tuning or geometry. In this paper, we clarified the relationship between these two perspectives. Some recent studies consider both tuning and geometry~\cite{Golden2016,Jazayeri2017,Zhou2018,Freeman2018,Chung2018,Tring2018,Ringach2019,Stringer2019,Sohn2019,Henaff2019,Minxha2020,Okazawa2021}. We hope this trend continues in the experimental literature and will be complemented by theoretical work toward a unified dual perspective on neural codes.

\section*{Competing interests}
The authors declare that they hold no competing interests.

\small{

}

\newpage
\section*{\LARGE Supplementary Information}

\subsection{Rotations, translations, and reflections conserve geometry and information content} \label{sec:rotinv}
Assume that $p(\mathbf{r}|\theta)$, the probability of response pattern ${\mathbf{r}}$ given stimulus $\theta$,  follows an isotropic multivariate Gaussian distribution $\mathcal{N}(\bf{f}(\theta),I_n)$. For a particular response $\mathbf{r}$, the likelihood function can be written as \begin{equation}\mathcal{L}(\theta|\mathbf{r}) \propto
e^{-\frac{1}{2}\|\mathbf{r}-\bf{f}(\theta)\|^2}
\end{equation}
Applying an arbitrary rotation using a rotation matrix $\mathbf{A}$ would lead to a new representation $\mathcal{N}(\mathbf{A}\cdot\bf{f}(\theta), I_n)$ with the same geometry. Because the noise is isotropic the rotation of the tuning curves rigidly rotates the entire joint probability function $p(\theta,\mathbf{r})$. For each original response pattern $\mathbf{r}$ generated by the original code, there is a new response pattern $\mathbf{r'}=\mathbf{A}\cdot \mathbf{r}$ that is equally probable. The likelihood function for the transformed code and response pattern can be expressed as
\begin{equation}\mathcal{L}(\theta|\mathbf{r}') \propto 
e^{-\frac{1}{2}\|\mathbf{A}\cdot (\mathbf{r}-\bf{f}(\theta))\|^2} = e^{-\frac{1}{2}\|\mathbf{r}-\bf{f}(\theta)\|^2}
\end{equation}
For every value of $\theta$, thus, $\mathcal{L}(\theta|\mathbf{r}')=\mathcal{L}(\theta|\mathbf{r})$. This conservation of the likelihood function holds not just for rotations, but also for translations and reflections: all transformations that conserve the geometry. Because the prior is also the same, these transformations have no influence on the performance of a Bayesian ideal observer. Choosing a different code with the same geometry yields the same estimation error distribution.

\begin{figure}
\centering
\includegraphics[keepaspectratio,width=.9\linewidth]{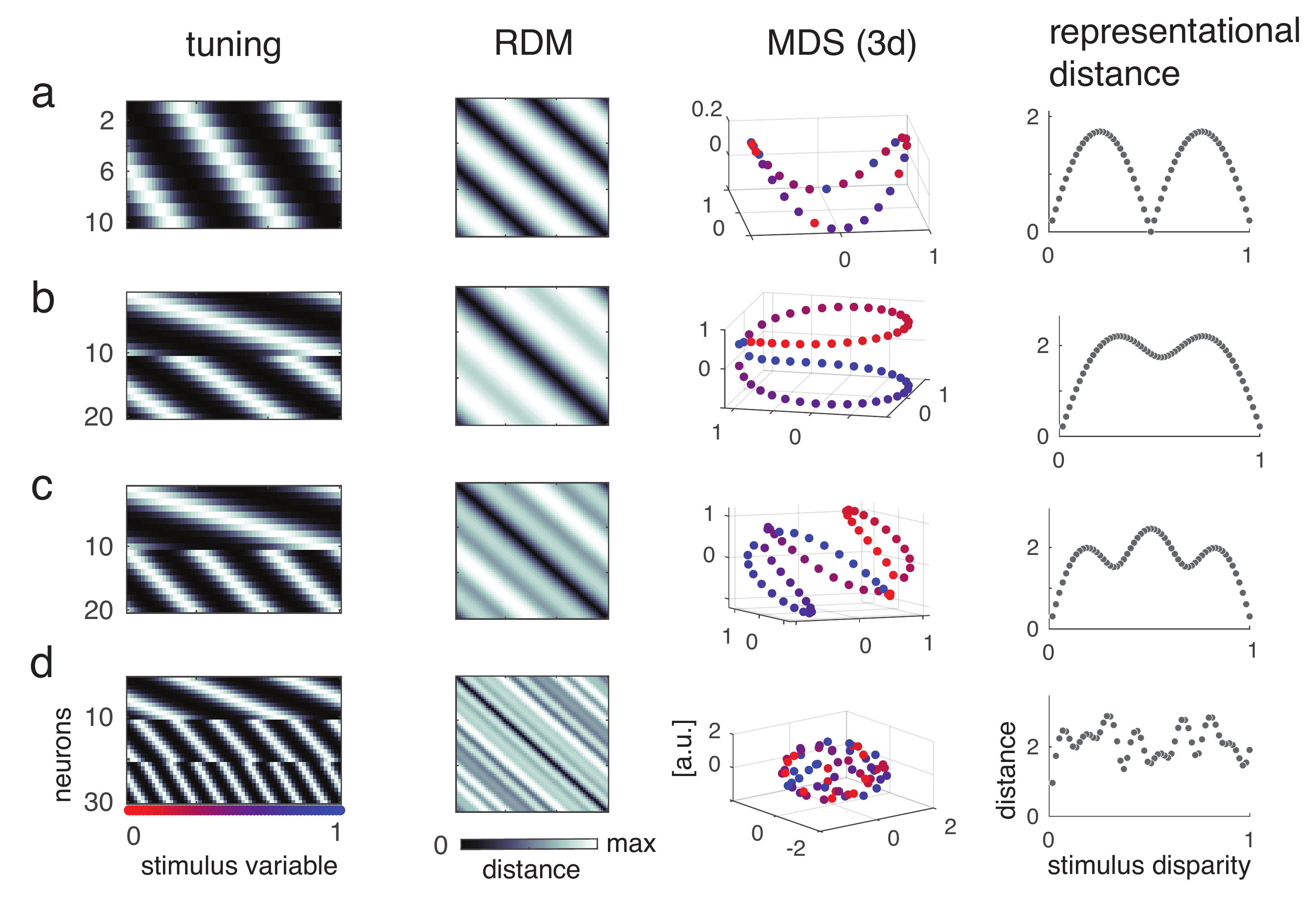}
\caption{\textbf{Encoding 1-d location with grid cells}. (\textbf{a}) A grid module for a single scale. (\textbf{b}) Two grid modules for two scales with a scale ratio of 2. (\textbf{c}) Two grid modules with a scale ratio of 3. (\textbf{d}) Three grid modules with scales [2, 5.5, 8.3]. }

\label{fig:app2}
\end{figure}

\subsection{An additional coding example: grid cells}
How can a population of neurons encode a position in an extended (and potentially unbounded) 1D or 2D space, such as the location of a rat in a corridor or 2D environment? For a finite environment, monotonic tuning functions could encode the coordinates. However, what if the rat keeps running once each neuron's activity has either saturated or gone to zero? It seems useful for neurons encoding the position in a large space to come on and off periodically. This principle is familiar from the way we write numbers (with ten digits advanced in a cycle) and from the Fourier transform. In both cases, we need multiple scales (digits of different significance and Fourier components of different frequencies) to disambiguate the code.

An encoding of 2D space is provided by grid cells. Originally discovered in rats~\cite{Fyhn2004,Hafting2005} and later in humans and other animal species~\cite{Fyhn2008,Doeller2010,Yartsev2011,Killian2012,Jacobs2013,Constantinescu2016,Bellmund2018}, grid cells provide a code of 2D position. A given grid cell will fire at locations clustered around the vertices of a triangular lattice over the environment. Different cells will fire around the vertices of triangular lattices of different phases and spatial scales~\cite{Hafting2005, Stensola2012}, collectively forming a multiscale code for location. To illustrate the representational geometry resulting from multiscale, periodic tuning functions, let us consider, for simplicity, a code for a 1D environment, such as a corridor (Fig.~\ref{fig:app2}).

We assume that each grid cell has a periodic tuning function, and different grid cells can have different phases and spatial scales. A population of grid cells with the same scale, where each neuron encodes a different phase, results in a cyclic representational geometry~\cite{Mcnaughton2006} (Fig.~\ref{fig:app2}a). As the animal keeps moving in the same direction, the population activity pattern traverses this cyclic path over and over. We can add another population, a second module, to encode a second scale. Two modules for different scales already result in a complex geometry that forms a cycle whose period is the least common multiple of the periods of the two modules (Fig.~\ref{fig:app2}b, c). The representational geometry is sensitive to the scale ratio between the different modules (compare Fig.~\ref{fig:app2} b and c, which have scale ratios of 2 and 3, respectively). For these codes, the representational distance is a smooth (albeit not a monotonic) function of the distance between two locations.

Fig.~\ref{fig:app2}d shows an example with 3 grid modules. The representation is highly distorted, and the representational distance is no longer a smooth function of the distance between two locations. Interestingly, most pairs of the points are well separated (Fig.~\ref{fig:app2}d, right), unlike the previous cases. This is due to the greater spread (higher entropy) of the response patterns (Fig.~\ref{fig:app2}d, third column), which increases the informational capacity of the code. As neighboring points along the manifold (corresponding to neighboring locations along the corridor) become more distinct in the representation, the FI increases. As pairs of points, in general, become more distinct, the MI increases. Maximizing the encoding capacity, thus will require spreading out the manifold to increase its length, corresponding to the total FI and, in case we desire high MI, to avoid its coming close to itself. There is likely a tradeoff between maximizing some notion of encoding capacity (e.g., FI or MI) and maintaining a simple (e.g., near monotonic) relationship between environmental and representational distance.


\begin{thebibliography}{255}
\bibitem{decharms2000neural} Christopher R DeCharms and Anthony Zador. Neural representation and the cortical code. \emph{Annual review of neuroscience}, 23(1):613--647, 2000.
\bibitem{kriegeskorte2019interpreting}Nikolaus Kriegeskorte and Pamela K Douglas. Interpreting encoding and decoding models. \emph{Current Opinion in Neurobiology,} 55:167--179, 2019.
\bibitem{Barlow1967}Horace B Barlow, Colin Blakemore, and John D Pettigrew. The neural mechanism of binocular depth discrimination. \emph{The Journal of physiology,} 193(2):327, 1967.
\bibitem{Campbell1968}FW Campbell, BG Cleland, GF Cooper, and Christina Enroth-Cugell. The angular selectivity of visual cortical cells to moving gratings. \emph{The Journal of Physiology,} 198(1):237--250, 1968.
\bibitem{Blakemore1972}Colin Blakemore, Adriana Fiorentini, and Lamberto Maffei. A second neural mechanism of binocular depth discrimination. \emph{The Journal of physiology,} 226(3):725--749, 1972.
\bibitem{Henry1974}Geoffrey H Henry, B Dreher, and PO Bishop. Orientation specificity of cells in cat striate cortex. \emph{Journal of Neurophysiology,} 37(6):1394--1409, 1974.
\bibitem{Hubel1959}David H Hubel and Torsten N Wiesel. Receptive fields of single neurones in the cat's striate cortex. \emph{The Journal of physiology,} 148(3):574--591, 1959.
\bibitem{Hubel1962}David H Hubel and Torsten N Wiesel. Receptive fields, binocular interaction and functional architecture in the cat's visual cortex. \emph{The Journal of physiology,} 160(1):106--154, 1962.
\bibitem{Hubel1968}David H Hubel and Torsten N Wiesel. Receptive fields and functional architecture of monkey striate cortex. \emph{The Journal of physiology,} 195(1):215--243, 1968.
\bibitem{Rose1974}David Rose and Colin Blakemore. An analysis of orientation selectivity in the cat's visual cortex. \emph{Experimental Brain Research,} 20(1):1--17, 1974.
\bibitem{Swindale1998}Nicholas V Swindale. Orientation tuning curves: empirical description and estimation of parameters. \emph{Biological cybernetics,} 78(1):45--56, 1998.
\bibitem{Georgopoulos1986}Apostolos P Georgopoulos, Andrew B Schwartz, and Ronald E Kettner. Neuronal population coding of movement direction. \emph{Science,} 233(4771):1416--1419, 1986.
\bibitem{Ben1995}Rani Ben-Yishai, R Lev Bar-Or, and Haim Sompolinsky. Theory of orientation tuning in visual cortex. \emph{Proceedings of the National Academy of Sciences,} 92(9):3844--3848, 1995.
\bibitem{Anderson2000}Jeffrey S Anderson, Ilan Lampl, Deda C Gillespie, and David Ferster. The contribution of noise to contrast invariance of orientation tuning in cat visual cortex. \emph{Science,} 290(5498):1968--1972, 2000.
\bibitem{Series2004}Peggy Seri{\`e}s, Peter E Latham, and Alexandre Pouget. Tuning curve sharpening for orientation selectivity: coding efficiency and the impact of correlations. \emph{Nature neuroscience,} 7(10):1129--1135, 2004.
\bibitem{Butts2006}Daniel A Butts and Mark S Goldman. Tuning curves, neuronal variability, and sensory coding. \emph{PLoS biology,} 4(4), 2006.
\bibitem{Campbell1969}FW Campbell, Go F Cooper, and Christina Enroth-Cugell. The spatial selectivity of the visual cells of the cat. \emph{The Journal of physiology,} 203(1):223--235, 1969.
\bibitem{Goldberg1969}Jay M Goldberg and Paul B Brown. Response of binaural neurons of dog superior olivary complex to dichotic tonal stimuli: some physiological mechanisms of sound localization. \emph{Journal of neurophysiology,} 32(4):613--636, 1969.
\bibitem{Suga1977}Nobuo Suga. Amplitude spectrum representation in the Doppler-shifted-CF processing area of the auditory cortex of the mustache bat. \emph{Science,} 196(4285):64--67, 1977.
\bibitem{o1978}John O'Keefe and Lynn Nadel. \emph{The hippocampus as a cognitive map.} Oxford: Clarendon Press, 1978.
\bibitem{Knudsen1982}Eric I Knudsen. Auditory and visual maps of space in the optic tectum of the owl. \emph{Journal of Neuroscience,} 2(9):1177--1194, 1982.
\bibitem{Maunsell1983}John H Maunsell and David C Van Essen. Functional properties of neurons in middle temporal visual area of the macaque monkey. I. Selectivity for stimulus direction, speed, and orientation. \emph{Journal of neurophysiology,} 49(5):1127--1147, 1983.
\bibitem{Taube1990}Jeffrey S Taube, Robert U Muller, and James B Ranck. Head-direction cells recorded from the postsubiculum in freely moving rats. I. Description and quantitative analysis. \emph{Journal of Neuroscience,} 10(2):420--435, 1990.
\bibitem{Deangelis1991}Gregory C DeAngelis, Izumi Ohzawa, and Ralph D Freeman. Depth is encoded in the visual cortex by a specialized receptive field structure. \emph{Nature,} 352(6331):156--159, 1991.
\bibitem{Johnson1992}Kenneth O Johnson and Steven S Hsiao. Neural mechanisms of tactual form and texture perception. \emph{Annual review of neuroscience,} 15(1):227--250, 1992.
\bibitem{Gallant1996}Jack L Gallant, Charles E Connor, Subrata Rakshit, James W Lewis, and David C Van Essen. Neural responses to polar, hyperbolic, and Cartesian gratings in area V4 of the macaque monkey. \emph{Journal of neurophysiology,} 76(4):2718--2739, 1996.
\bibitem{Pasupathy2002}Anitha Pasupathy and Charles E Connor. Population coding of shape in area V4. \emph{Nature neuroscience,} 5(12):1332--1338, 2002.
\bibitem{Nieder2002}Andreas Nieder, David J Freedman, and Earl K Miller. Representation of the quantity of visual items in the primate prefrontal cortex. \emph{Science,} 297(5587):1708--1711, 2002.
\bibitem{Fyhn2004}Marianne Fyhn, Sturla Molden, Menno P Witter, Edvard I Moser, and May-Britt Moser. Spatial representation in the entorhinal cortex. \emph{Science,} 305(5688):1258--1264, 2004.
\bibitem{Hafting2005}Torkel Hafting, Marianne Fyhn, Sturla Molden, May-Britt Moser, and Edvard I Moser. Microstructure of a spatial map in the entorhinal cortex. \emph{Nature,} 436(7052):801, 2005.
\bibitem{Young1992}Malcolm P Young and Shigeru Yamane. Sparse population coding of faces in the inferotemporal cortex. \emph{Science,} 256(5061):1327--1331, 1992.
\bibitem{Tsao2006}Doris Y Tsao, Winrich A Freiwald, Roger BH Tootell, and Margaret S Livingstone. A cortical region consisting entirely of face-selective cells. \emph{Science,} 311(5761):670--674, 2006.
\bibitem{Rigotti2013}Mattia Rigotti, Omri Barak, Melissa R Warden, Xiao-Jing Wang, Nathaniel D Daw, Earl K Miller, and Stefano Fusi. The importance of mixed selectivity in complex cognitive tasks. \emph{Nature,} 497(7451):585--590, 2013.
\bibitem{Chang2017}Le Chang and Doris Y Tsao. The code for facial identity in the primate brain. \emph{Cell,} 169(6):1013--1028, 2017.
\bibitem{Maffei1976}Lamberto Maffei and Adriana Fiorentini. The unresponsive regions of visual cortical receptive fields. \emph{Vision research,} 16(10):1131--IN5, 1976.
\bibitem{Gilbert1990}Charles D Gilbert and Torsten N Wiesel. The influence of contextual stimuli on the orientation selectivity of cells in primary visual cortex of the cat. \emph{Vision research,} 30(11):1689--1701, 1990.
\bibitem{Knierim1992}James J Knierim and David C Van Essen. Neuronal responses to static texture patterns in area V1 of the alert macaque monkey. \emph{Journal of neurophysiology,} 67(4):961--980, 1992.
\bibitem{Maffei1973}L Maffei, A Fiorentini, and S Bisti. Neural correlate of perceptual adaptation to gratings. \emph{Science,} 182(4116):1036--1038, 1973.
\bibitem{Movshon1979}J Anthony Movshon and Peter Lennie. Pattern-selective adaptation in visual cortical neurones. \emph{Nature,} 278(5707):850, 1979.
\bibitem{Dragoi2000}Valentin Dragoi, Jitendra Sharma, and Mriganka Sur. Adaptation-induced plasticity of orientation tuning in adult visual cortex. \emph{Neuron,} 28(1):287--298, 2000.
\bibitem{Benucci2013}Andrea Benucci, Aman B Saleem, and Matteo Carandini. Adaptation maintains population homeostasis in primary visual cortex. \emph{Nature neuroscience,} 16(6):724, 2013.
\bibitem{Dean2005}Isabel Dean, Nicol S Harper, and David McAlpine. Neural population coding of sound level adapts to stimulus statistics. \emph{Nature neuroscience,} 8(12):1684, 2005.
\bibitem{Ulanovsky2004}Nachum Ulanovsky, Liora Las, Dina Farkas, and Israel Nelken. Multiple time scales of adaptation in auditory cortex neurons. \emph{Journal of Neuroscience,} 24(46):10440--10453, 2004.
\bibitem{Grill2006}Kalanit Grill-Spector, Richard Henson, and Alex Martin. Repetition and the brain: neural models of stimulus-specific effects. \emph{Trends in cognitive sciences,} 10(1):14--23, 2006.
\bibitem{Solomon2014}Samuel G Solomon and Adam Kohn. Moving sensory adaptation beyond suppressive effects in single neurons. \emph{Current Biology,} 24(20):R1012--R1022, 2014.
\bibitem{Alink2018}Arjen Alink, Hunar Abdulrahman, and Richard N Henson. Forward models demonstrate that repetition suppression is best modelled by local neural scaling. \emph{Nature communications,} 9(1):1--10, 2018.
\bibitem{Treue1999}Stefan Treue and Julio C Martinez Trujillo. Feature-based attention influences motion processing gain in macaque visual cortex. \emph{Nature,} 399(6736):575, 1999.
\bibitem{McAdams1999}Carrie J McAdams and John HR Maunsell. Effects of attention on orientation-tuning functions of single neurons in macaque cortical area V4. \emph{Journal of Neuroscience,} 19(1):431--441, 1999.
\bibitem{Reynolds2000}John H Reynolds, Tatiana Pasternak, and Robert Desimone. Attention increases sensitivity of V4 neurons. \emph{Neuron,} 26(3):703--714, 2000.
\bibitem{Schoups2001}Aniek Schoups, Rufin Vogels, Ning Qian, and Guy Orban. Practising orientation identification improves orientation coding in V1 neurons. \emph{Nature,} 412(6846):549--553, 2001.
\bibitem{Ghose2002}Geoffrey M Ghose, Tianming Yang, and John HR Maunsell. Physiological correlates of perceptual learning in monkey V1 and V2. \emph{Journal of neurophysiology,} 87(4):1867--1888, 2002.
\bibitem{Crist2001}Roy E Crist, Wu Li, and Charles D Gilbert. Learning to see: experience and attention in primary visual cortex. \emph{Nature neuroscience,} 4(5):519--525, 2001.
\bibitem{Churchland2012}Mark M Churchland, John P Cunningham, Matthew T Kaufman, Justin D Foster, Paul Nuyujukian, Stephen I Ryu, and Krishna V Shenoy. Neural population dynamics during reaching. \emph{Nature,} 487(7405):51--56, 2012.
\bibitem{Shenoy2013}Krishna V Shenoy, Maneesh Sahani, and Mark M Churchland. Cortical control of arm movements: a dynamical systems perspective. \emph{Annual review of neuroscience,} 36:337--359, 2013.
\bibitem{Churchland2011}Anne K Churchland, Roozbeh Kiani, Rishidev Chaudhuri, Xiao-Jing Wang, Alexandre Pouget, and Michael N Shadlen. Variance as a signature of neural computations during decision making. \emph{Neuron,} 69(4):818--831, 2011.
\bibitem{Buzsaki2004}Gy{\"o}rgy Buzs{\'a}ki. Large-scale recording of neuronal ensembles. \emph{Nature neuroscience,} 7(5):446--451, 2004.
\bibitem{Stevenson2011}Ian H Stevenson and Konrad P Kording. How advances in neural recording affect data analysis. \emph{Nature neuroscience,} 14(2):139--142, 2011.
\bibitem{Jun2017}James J Jun, Nicholas A Steinmetz, Joshua H Siegle, Daniel J Denman, Marius Bauza, Brian Barbarits, Albert K Lee, Costas A Anastassiou, Alexandru Andrei, {\c{C}}a{\u{g}}atay Ayd{\i}n, et al. Fully integrated silicon probes for high-density recording of neural activity. \emph{Nature,} 551(7679):232--236, 2017.
\bibitem{Biswal1995}Bharat Biswal, F Zerrin Yetkin, Victor M Haughton, and James S Hyde. Functional connectivity in the motor cortex of resting human brain using echo-planar MRI. \emph{Magnetic resonance in medicine,} 34(4):537--541, 1995.
\bibitem{Fox2007}Michael D Fox and Marcus E Raichle. Spontaneous fluctuations in brain activity observed with functional magnetic resonance imaging. \emph{Nature reviews neuroscience,} 8(9):700--711, 2007.
\bibitem{Shepard1970}Roger N Shepard and Susan Chipman. Second-order isomorphism of internal representations: Shapes of states. \emph{Cognitive psychology,} 1(1):1--17, 1970.
\bibitem{Edelman1998a}Shimon Edelman, Kalanit Grill-Spector, Tamar Kushnir, and Rafael Malach. Toward direct visualization of the internal shape representation space by fMRI. \emph{Psychobiology,} 26(4):309--321, 1998.
\bibitem{Edelman1998b}Shimon Edelman. Representation is representation of similarities. \emph{The Behavioral and brain sciences,} 21(4):449, 1998.
\bibitem{Norman2006}Kenneth A Norman, Sean M Polyn, Greg J Detre, and James V Haxby. Beyond mind-reading: multi-voxel pattern analysis of fMRI data. \emph{Trends in cognitive sciences,} 10(9):424--430, 2006.
\bibitem{Diedrichsen2017}J{\"o}rn Diedrichsen and Nikolaus Kriegeskorte. Representational models: A common framework for understanding encoding, pattern-component, and representational-similarity analysis. \emph{PLoS computational biology,} 13(4):e1005508, 2017.
\bibitem{Kriegeskorte2008a}Nikolaus Kriegeskorte, Marieke Mur, and Peter Bandettini. Representational similarity analysis-connecting the branches of systems neuroscience. \emph{Frontiers in systems neuroscience,} 2:4, 2008.
\bibitem{Kriegeskorte2008b}Nikolaus Kriegeskorte, Marieke Mur, Douglas A Ruff, Roozbeh Kiani, Jerzy Bodurka, Hossein Esteky, Keiji Tanaka, and Peter A Bandettini. Matching categorical object representations in inferior temporal cortex of man and monkey. \emph{Neuron,} 60(6):1126--1141, 2008.
\bibitem{Connolly2012}Andrew C Connolly, J Swaroop Guntapalli, Jason Gors, Michael Hanke, Yaroslav O Halchenko, Yu-Chien Wu, Herv{\'e} Abdi, and James V Haxby. The representation of biological classes in the human brain. \emph{Journal of Neuroscience,} 32(8):2608--2618, 2012.
\bibitem{Xue2010}Gui Xue, Qi Dong, Chuansheng Chen, Zhonglin Lu, Jeannette A Mumford, and Russell A Poldrack. Greater neural pattern similarity across repetitions is associated with better memory. \emph{Science,} 330(6000):97--101, 2010.
\bibitem{Khaligh2014}Seyed-Mahdi Khaligh-Razavi and Nikolaus Kriegeskorte. Deep supervised, but not unsupervised, models may explain IT cortical representation. \emph{PLoS computational biology,} 10(11):e1003915, 2014.
\bibitem{Yamins2014}Daniel LK Yamins, Ha Hong, Charles F Cadieu, Ethan A Solomon, Darren Seibert, and James J DiCarlo. Performance-optimized hierarchical models predict neural responses in higher visual cortex. \emph{Proceedings of the National Academy of Sciences,} 111(23):8619--8624, 2014.
\bibitem{Cichy2014}Radoslaw Martin Cichy, Dimitrios Pantazis, and Aude Oliva. Resolving human object recognition in space and time. \emph{Nature neuroscience,} 17(3):455, 2014.
\bibitem{Freeman2018}Jonathan B Freeman, Ryan M Stolier, Jeffrey A Brooks, and Benjamin S Stillerman. The neural representational geometry of social perception. \emph{Current opinion in psychology,} 24:83--91, 2018.
\bibitem{Kietzmann2019}Tim C Kietzmann, Courtney J Spoerer, Lynn KA S{\"o}rensen, Radoslaw M Cichy, Olaf Hauk, and Nikolaus Kriegeskorte. Recurrence is required to capture the representational dynamics of the human visual system. \emph{Proceedings of the National Academy of Sciences,} 116(43):21854--21863, 2019.
\bibitem{Kriegeskorte2019}Nikolaus Kriegeskorte and J{\"o}rn Diedrichsen. Peeling the Onion of Brain Representations. \emph{Annual review of neuroscience,} 42:407--432, 2019.
\bibitem{Kriegeskorte2008}Nikolaus Kriegeskorte, Marieke Mur, Douglas A Ruff, Roozbeh Kiani, Jerzy Bodurka, Hossein Esteky, Keiji Tanaka, and Peter A Bandettini. Matching categorical object representations in inferior temporal cortex of man and monkey. \emph{Neuron,} 60(6):1126--1141, 2008.
\bibitem{Nili2014}Hamed Nili, Cai Wingfield, Alexander Walther, Li Su, William Marslen-Wilson, and Nikolaus Kriegeskorte. A toolbox for representational similarity analysis. \emph{PLoS computational biology,} 10(4):e1003553, 2014.
\bibitem{Kriegeskorte2016}Nikolaus Kriegeskorte and J{\"o}rn Diedrichsen. Inferring brain-computational mechanisms with models of activity measurements. \emph{Philosophical Transactions of the Royal Society B: Biological Sciences,} 371(1705):20160278, 2016.
\bibitem{Kriegeskorte2013}Nikolaus Kriegeskorte and Rogier A Kievit. Representational geometry: integrating cognition, computation, and the brain. \emph{Trends in cognitive sciences,} 17(8):401--412, 2013.
\bibitem{Dumoulin2008}Serge O Dumoulin and Brian A Wandell. Population receptive field estimates in human visual cortex. \emph{Neuroimage,} 39(2):647--660, 2008.
\bibitem{Kay2008}Kendrick N Kay, Thomas Naselaris, Ryan J Prenger, and Jack L Gallant. Identifying natural images from human brain activity. \emph{Nature,} 452(7185):352--355, 2008.
\bibitem{Naselaris2009}Thomas Naselaris, Ryan J Prenger, Kendrick N Kay, Michael Oliver, and Jack L Gallant. Bayesian reconstruction of natural images from human brain activity. \emph{Neuron,} 63(6):902--915, 2009.
\bibitem{Freiwald2010}Winrich A Tsao and Doris Y Tsao. Functional compartmentalization and viewpoint generalization within the macaque face-processing system. \emph{Science,} 330(6005):845--851, 2010.
\bibitem{Ringach2019}Dario L Ringach. The geometry of masking in neural populations. \emph{Nature communications,} 10(1):1--11, 2019.
\bibitem{Fisher1922}Ronald A Fisher. On the mathematical foundations of theoretical statistics. \emph{Philosophical Transactions of the Royal Society of London. Series A, Containing Papers of a Mathematical or Physical Character,} 222(594-604):309--368, 1922.
\bibitem{Shannon1948}Claude Elwood Shannon. A mathematical theory of communication. \emph{Bell system technical journal,} 27(3):379--423, 1948.
\bibitem{van1997}Rob R de Ruyter van Steveninck, Geoffrey D Lewen, Steven P Strong, Roland Koberle, and William Bialek. Reproducibility and variability in neural spike trains. \emph{Science,} 275(5307):1805--1808, 1997.
\bibitem{Rieke1999}Fred Rieke, David Warland, Rob de Ruyter Van Steveninck, Bialek, William S, et al. Spikes: exploring the neural code. MIT Press Cambridge, 1999.
\bibitem{Borst1999}Alexander Thorst and Fr{\'e}d{\'e}ric E Theunissen. Information theory and neural coding. \emph{Nature neuroscience,} 2(11):947, 1999.
\bibitem{Fairhall2001}Adrienne L Fairhall, Geoffrey D Lewen, William Bialek, and Robert R de Ruyter van Steveninck. Efficiency and ambiguity in an adaptive neural code. \emph{Nature,} 412(6849):787, 2001.
\bibitem{Theunissen1991}Fr{\'e}d{\'e}ric E Theunissen and John P Miller. Representation of sensory information in the cricket cercal sensory system. II. Information theoretic calculation of system accuracy and optimal tuning-curve widths of four primary interneurons. \emph{Journal of neurophysiology,} 66(5):1690--1703, 1991.
\bibitem{Roddey1996}J Cooper Roddey and Gwen A Jacobs. Information theoretic analysis of dynamical encoding by filiform mechanoreceptors in the cricket cercal system. \emph{Journal of neurophysiology,} 75(4):1365--1376, 1996.
\bibitem{Theunissen1996}Fr{\'e}d{\'e}ric Theunissen, J Cooper Roddey, Steven Stufflebeam, Heather Clague, and JP Miller. Information theoretic analysis of dynamical encoding by four identified primary sensory interneurons in the cricket cercal system. \emph{Journal of Neurophysiology,} 75(4):1345--1364, 1996.
\bibitem{Brenner2000}Naama Brenner, William Bialek, and Rob de Ruyter Van Steveninck. Adaptive rescaling maximizes information transmission. \emph{Neuron,} 26(3):695--702, 2000.
\bibitem{Ganguli2014}Deep Ganguli and Eero P Simoncelli. Efficient sensory encoding and Bayesian inference with heterogeneous neural populations. \emph{Neural computation,} 26(10):2103--2134, 2014.
\bibitem{Wei2015}Xue-Xin Wei and Alan A Stocker. A Bayesian observer model constrained by efficient coding can explain 'anti-Bayesian' percepts. \emph{Nature neuroscience,} 18(10):1509, 2015.
\bibitem{Wei2012}Xue-Xin Wei and Alan A Stocker. Efficient coding provides a direct link between prior and likelihood in perceptual Bayesian inference. In \emph{Advances in neural information processing systems,} pages 1304--1312, 2012.
\bibitem{Mcdonnell2008}Mark D McDonnell and Nigel G Stocks. Maximally informative stimuli and tuning curves for sigmoidal rate-coding neurons and populations. \emph{Physical review letters,} 101(5):058103, 2008.
\bibitem{Barlow1961}Horace B. Barlow et al. Possible principles underlying the transformation of sensory messages. \emph{Sensory communication,} 1:217--234, 1961.
\bibitem{Linsker1988}Ralph Linsker. Self-organization in a perceptual network. \emph{Computer,} 21(3):105--117, 1988.
\bibitem{laughlin1981}Simon Laughlin. A simple coding procedure enhances a neuron's information capacity. \emph{Zeitschrift f{\"u}r Naturforschung c,} 36(9-10):910--912, 1981.
\bibitem{van1992}Johannes H van Hateren. A theory of maximizing sensory information. \emph{Biological cybernetics,} 68(1):23--29, 1992.
\bibitem{Atick1992}Joseph J Atick. Could information theory provide an ecological theory of sensory processing? \emph{Network: Computation in neural systems,} 3(2):213--251, 1992.
\bibitem{Dong1995}Dawei W Dong and Joseph J Atick. Statistics of natural time-varying images. \emph{Network: Computation in Neural Systems,} 6(3):345--358, 1995.
\bibitem{Olshausen1996}Bruno A Olshausen and David J Field. Emergence of simple-cell receptive field properties by learning a sparse code for natural images. \emph{Nature,} 381(6583):607, 1996.
\bibitem{Bell1997}Anthony J Bell and Terrence J Sejnowski. The “independent components” of natural scenes are edge filters. \emph{Vision research,} 37(23):3327--3338, 1997.
\bibitem{Simoncelli2001}Eero P Simoncelli and Bruno A Olshausen. Natural image statistics and neural representation. \emph{Annual review of neuroscience,} 24(1):1193--1216, 2001.
\bibitem{Ganguli2010}Deep Ganguli and Eero P Simoncelli. Implicit encoding of prior probabilities in optimal neural populations. In \emph{Advances in neural information processing systems,} pages 658--666, 2010.
\bibitem{Wei2016}Xue-Xin Wei and Alan A Stocker. Mutual information, Fisher information, and efficient coding. \emph{Neural computation,} 28(2):305--326, 2016.
\bibitem{Quiroga2009}Rodrigo Quian Quiroga and Stefano Panzeri. Extracting information from neuronal populations: information theory and decoding approaches. \emph{Nature Reviews Neuroscience,} 10(3):173, 2009.
\bibitem{deweese1999}Michael R DeWeese and Markus Meister. How to measure the information gained from one symbol. \emph{Network: Computation in Neural Systems,} 10(4):325--340, 1999.
\bibitem{Butts2003}Daniel A Butts. How much information is associated with a particular stimulus? \emph{Network: Computation in Neural Systems,} 14(2):177--187, 2003.
\bibitem{Montgomery2010}Nathan Montgomery and Michael Wehr. Auditory cortical neurons convey maximal stimulus-specific information at their best frequency. \emph{Journal of Neuroscience,} 30(40):13362--13366, 2010.
\bibitem{Lehmann2006}Erich L Lehmann and George Casella. \emph{Theory of point estimation.} Springer Science \& Business Media, 2006.
\bibitem{Harper2004}Nicol S Harper and David McAlpine. Optimal neural population coding of an auditory spatial cue. \emph{Nature,} 430(7000):682, 2004.
\bibitem{Gutnisky2008}Diego A Gutnisky and Valentin Dragoi. Adaptive coding of visual information in neural populations. \emph{Nature,} 452(7184):220, 2008.
\bibitem{Brunel1998}Nicolas Brunel and Jean-Pierre Nadal. Mutual information, Fisher information, and population coding. \emph{Neural computation,} 10(7):1731--1757, 1998.
\bibitem{Zhang1999}Kechen Zhang and Terrence J Sejnowski. Neuronal tuning: To sharpen or broaden? \emph{Neural computation,} 11(1):75--84, 1999.
\bibitem{Pouget1999}Alexandre Pouget, Sophie Deneve, Jean-Christophe Ducom, and Peter E Latham. Narrow versus wide tuning curves: What's best for a population code? \emph{Neural computation,} 11(1):85--90, 1999.
\bibitem{Durant2007}Szonya Durant, Colin WG Clifford, Nathan A Crowder, Nicholas SC Price, and Michael R Ibbotson. Characterizing contrast adaptation in a population of cat primary visual cortical neurons using Fisher information. \emph{JOSA A,} 24(6):1529--1537, 2007.
\bibitem{Ecker2011}Alexander S Ecker, Philipp Berens, Andreas S Tolias, and Matthias Bethge. The effect of noise correlations in populations of diversely tuned neurons. \emph{Journal of Neuroscience,} 31(40):14272--14283, 2011.
\bibitem{Yarrow2012}Stuart Yarrow, Edward Challis, and Peggy Seri{\`e}s. Fisher and Shannon information in finite neural populations. \emph{Neural computation,} 24(7):1740--1780, 2012.
\bibitem{Lin2015}I-Chun Lin, Michael Okun, Matteo Carandini, and Kenneth D Harris. The nature of shared cortical variability. \emph{Neuron,} 87(3):644--656, 2015.
\bibitem{Arandia2016}I{\~n}igo Arandia-Romero, Seiji Tanabe, Jan Drugowitsch, Adam Kohn, and Rub{\'e}n Moreno-Bote. Multiplicative and additive modulation of neuronal tuning with population activity affects encoded information. \emph{Neuron,} 89(6):1305--1316, 2016.
\bibitem{Zohary1994}Ehud Zohary, Michael N Shadlen, and William T Newsome. Correlated neuronal discharge rate and its implications for psychophysical performance. \emph{Nature,} 370(6485):140--143, 1994.
\bibitem{Shadlen1996}Michael N Shadlen, Kenneth H Britten, William T Newsome, and J Anthony Movshon. A computational analysis of the relationship between neuronal and behavioral responses to visual motion. \emph{Journal of Neuroscience,} 16(4):1486--1510, 1996.
\bibitem{Bair2001}Wyeth Bair, Ehud Zohary, and William T Newsome. Correlated firing in macaque visual area MT: time scales and relationship to behavior. \emph{Journal of Neuroscience,} 21(5):1676--1697, 2014.
\bibitem{Kohn2005}Adam Kohn and Matthew A Smith. Stimulus dependence of neuronal correlation in primary visual cortex of the macaque. \emph{Journal of Neuroscience,} 25(14):3661--3673, 2005.
\bibitem{Cohen2009}Marlene R Cohen and John HR Maumsell. Attention improves performance primarily by reducing interneuronal correlations. \emph{Nature neuroscience,} 12(12):1594, 2009.
\bibitem{Cohen2011}Marlene R Cohen and Adam Kohn. Measuring and interpreting neuronal correlations. \emph{Nature neuroscience,} 14(7):811, 2011.
\bibitem{Abbott1999}Larry F Abbott and Peter Dayan. The effect of correlated variability on the accuracy of a population code. \emph{Neural computation,} 11(1):91--101, 1999.
\bibitem{Yoon1999effect}Hyoungsoo Yoon and Haim Sompolinsky. The effect of correlations on the Fisher information of population codes. In \emph{Advances in neural information processing systems,} pages 167--173, 1999.
\bibitem{Nirenberg2003}Sheila Nirenberg and Peter E Latham. Decoding neuronal spike trains: How important are correlations? \emph{Proceedings of the National Academy of Sciences,} 100(12):7348--7353, 2003.
\bibitem{Latham2005}Peter E Latham and Sheila Nirenberg. Synergy, redundancy, and independence in population codes, revisited. \emph{Journal of Neuroscience,} 25(21):5195--5206, 2005.
\bibitem{Pola2003}G Pola, A Thiele, KP Hoffmann, and S Panzeri. An exact method to quantify the information transmitted by different mechanisms of correlational coding. \emph{Network-Computation in Neural Systems,} 14(1):35--60, 2003.
\bibitem{Moreno2014}Rub{\'e}n Moreno-Bote, Jeffrey Beck, Ingmar Kanitscheider, Xaq Pitkow, Peter Latham, and Alexandre Pouget. Information-limiting correlations. \emph{Nature neuroscience,} 17(10):1410, 2014.
\bibitem{Averbeck2006}Bruno B Averbeck, Peter E Latham, and Alexandre Pouget. Neural correlations, population coding and computation. \emph{Nature reviews neuroscience,} 7(5):358, 2006.
\bibitem{Kohn2016}Adam Kohn, Ruben Coen-Cagli, Ingmar Kanitscheider, and Alexandre Pouget. Correlations and neuronal population information. \emph{Annual review of neuroscience,} 39:237--256, 2016.
\bibitem{Bartlett1936}MSo Bartlett. The square root transformation in analysis of variance. \emph{Supplement to the Journal of the Royal Statistical Society,} 3(1):68--78, 1936.
\bibitem{Anscombe1948}Francis J Anscombe. The transformation of Poisson, binomial and negative-binomial data. \emph{Biometrika,} 35(3/4):246--254, 1948.
\bibitem{Seung2000}H Sebastian Seung and Daniel D Lee. The manifold ways of perception. \emph{Science,} 290(5500):2268--2269, 2000.
\bibitem{Dicarlo2007}James J DiCarlo and David D Cox. Untangling invariant object recognition. \emph{Trends in cognitive sciences,} 11(8):333--341, 2007.
\bibitem{Jazayeri2017}Mehrdad Jazayeri and Arash Afraz. Navigating the neural space in search of the neural code. \emph{Neuron,} 93(5):1003--1014, 2017.
\bibitem{Zhou2018}Yuansheng Zhou, Brian H Smith, and Tatyana O Sharpee. Hyperbolic geometry of the olfactory space. \emph{Science advances,} 4(8):eaaq1458, 2018.
\bibitem{Stringer2019}Carsen Stringer, Marius Pachitariu, Nicholas Steinmetz, Matteo Carandini, and Kenneth D Harris. High-dimensional geometry of population responses in visual cortex. \emph{Nature,} page 1, 2019.
\bibitem{Bartlett1947}Maurice S Bartlett. The use of transformations. \emph{Biometrics,} 3(1):39--52, 1947.
\bibitem{Walther2016}Alexander Walther, Hamed Nili, Naveed Ejaz, Arjen Alink, Nikolaus Kriegeskorte, and J{\"o}rn Diedrichsen. Reliability of dissimilarity measures for multi-voxel pattern analysis. \emph{Neuroimage,} 137:188--200, 2016.
\bibitem{Green1966}David Marvin Green, John A Swets, et al. \emph{Signal detection theory and psychophysics,} volume 1. Wiley New York, 1966.
\bibitem{Knill1996}David C Knill and Whitman Richards. \emph{Perception as Bayesian inference.} Cambridge University Press, 1996.
\bibitem{Geisler2011}Wilson S Geisler. Contributions of ideal observer theory to vision research. \emph{Vision research,} 51(7):771--781, 2011.
\bibitem{Tomassini2010}Alessandro Tomassini, Michael J Morgan, and Joshua A Solomon. Orientation uncertainty reduces perceived obliquity. \emph{Vision research,} 50(5):541--547, 2010.
\bibitem{Girshick2011}Ahna R Girshick, Michael S Landy, and Eero P Simoncelli. Cardinal rules: visual orientation perception reflects knowledge of environmental statistics. \emph{Nature neuroscience,} 14(7):926, 2011.
\bibitem{Ruben2015}Ruben S Van Bergen, Wei Ji Ma, Michael S Pratte, and Janneke FM Jehee. Sensory uncertainty decoded from visual cortex predicts behavior. \emph{Nature Neuroscience,} 18(12):1728, 2015.
\bibitem{Sanger1996}T David Sanger. Probability density estimation for the interpretation of neural population codes. \emph{Journal of neurophysiology,} 76(4):2790--2793, 1996.
\bibitem{Zhang1998}Kechen Zhang, Iris Ginzburg, Bruce L McNaughton, and Terrence J Sejnowski. Interpreting neuronal population activity by reconstruction: unified framework with application to hippocampal place cells. \emph{Journal of neurophysiology,} 79(2):1017--1044, 1998.
\bibitem{Oram1998}Mike W Oram, Peter F{\"o}ldi{\'a}k, David I Perrett, and Frank Sengpiel. The 'Ideal Homunculus': decoding neural population signals. \emph{Trends in neurosciences,} 21(6):259--265, 1998.
\bibitem{Series2009}Peggy Seri{\`e}s, Alan A Stocker, and Eero P Simoncelli. Is the homunculus “aware” of sensory adaptation? \emph{Neural Computation,} 21(12):3271--3304, 2009.
\bibitem{Gu2010}Yong Gu, Christopher R Fetsch, Babatunde Adeyemo, Gregory C DeAngelis, and Dora E Angelaki. Decoding of MSTd population activity accounts for variations in the precision of heading perception. \emph{Neuron,} 66(4):596--609, 2010.
\bibitem{Graf2011}Arnulf BA Graf, Adam Kohn, Mehrdad Jazayeri, and J Anthony Movshon. Decoding the activity of neuronal populations in macaque primary visual cortex. \emph{Nature neuroscience,} 14(2):239, 2011.
\bibitem{Bays2014}Paul M Bays. Noise in neural populations accounts for errors in working memory. \emph{Journal of Neuroscience,} 34(10):3632--3645, 2014.
\bibitem{Zhang2008}Weiwei Zhang and Steven J Luck. Discrete fixed-resolution representations in visual working memory. \emph{Nature,} 453(7192):233, 2008.
\bibitem{Ma2014}Wei Ji Ma, Masud Husain, and Paul M Bays. Changing concepts of working memory. \emph{Nature neuroscience,} 17(3):347, 2014.
\bibitem{Levy1996}William B Levy and Robert A Baxter. Energy efficient neural codes. \emph{Neural computation,} 8(3):531--543, 1996.
\bibitem{Laughlin2001}Simon B Laughlin. Energy as a constraint on the coding and processing of sensory information. \emph{Current opinion in neurobiology,} 11(4):475--480, 2001.
\bibitem{Balasubramanian2002}Vijay Balasubramanian and Michael J Berry. A test of metabolically efficient coding in the retina. \emph{Network: Computation in Neural Systems,} 13(4):531--552, 2002.
\bibitem{Seung1993}H Sebastian Seung and Haim Sompolinsky. Simple models for reading neuronal population codes. \emph{Proceedings of the National Academy of Sciences,} 90(22):10749--10753, 1993.
\bibitem{Wei2017}Xue-Xin Wei and Alan A Stocker. Lawful relation between perceptual bias and discriminability. \emph{Proceedings of the National Academy of Sciences,} 114(38):10244--10249, 2017.
\bibitem{Clarke1994}Bertrand S Clarke and Andrew R Barron. Jeffreys' prior is asymptotically least favorable under entropy risk. \emph{Journal of Statistical Planning and Inference,} 41(1):37--60, 1994.
\bibitem{Appelle1972}Stuart Appelle. Perception and discrimination as a function of stimulus orientation: the" oblique effect" in man and animals. \emph{Psychological bulletin,} 78(4):266, 1972.
\bibitem{Li2018}Vickie Li, Elizabeth Michael, Jan Balaguer, Santiago Herce Casta{\~n}{\'o}n, and Christopher Summerfield. \emph{Proceedings of the National Academy of Sciences,} 115(38):E8825--E8834, 2018.
\bibitem{Attneave1954}Fred Attneave. Some informational aspects of visual perception. \emph{Psychological review,} 61(3):183, 1954.
\bibitem{Hateren1998}J Hans Van Hateren and Arjen van der Schaaf. Independent component filters of natural images compared with simple cells in primary visual cortex. \emph{Proceedings of the Royal Society of London. Series B: Biological Sciences,} 265(1394):359--366, 1998.
\bibitem{Schwartz2001}Odelia Schwartz and Eero P Simoncelli. Natural signal statistics and sensory gain control. \emph{Nature neuroscience,} 4(8):819, 2001.
\bibitem{Lewicki2002}Michael S Lewicki. Efficient coding of natural sounds. \emph{Nature neuroscience,} 5(4):356, 2002.
\bibitem{Wei2015grid}Xue-Xin Wei, Jason Prentice, and Vijay Balasubramanian. A principle of economy predicts the functional architecture of grid cells. \emph{Elife,} 4:e08362, 2015.
\bibitem{Mosheiff2017}Noga Mosheiff, Haggai Agmon, Avraham Moriel, and Yoram Burak. An efficient coding theory for a dynamic trajectory predicts non-uniform allocation of entorhinal grid cells to modules. \emph{PLoS computational biology,} 13(6):e1005597, 2017.
\bibitem{Polania2019}Rafael Polania, Michael Woodford, and Christian C Ruff. Efficient coding of subjective value. \emph{Nature neuroscience,} 22(1):134, 2019.
\bibitem{de1982}Russell L De Valois, E William Yund, and Norva Hepler. The orientation and direction selectivity of cells in macaque visual cortex. \emph{Vision research,} 22(5):531--544, 1982.
\bibitem{Nover2005}Harris Nover, Charles H Anderson, and Gregory C DeAngelis. A logarithmic, scale-invariant representation of speed in macaque middle temporal area accounts for speed discrimination performance. \emph{Journal of Neuroscience,} 25(43):10049--10060, 2005.
\bibitem{Stocker2009b}Alan A Stocker and Eero P Simoncelli. Visual motion aftereffects arise from a cascade of two isomorphic adaptation mechanisms. \emph{Journal of Vision,} 9(9):9, 2009.
\bibitem{Harvey2012} Christopher D Harvey, Philip Coen, and David W Tank. Choice-specific sequences in parietal cortex during a virtual-navigation decision task.\emph{Nature,} 484 (7392) 62--68, 2012.
\bibitem{Lebedev2019}Mikhail A Lebedev, Alexei Ossadtchi, Nil Adell Mill, N{\'u}ria Armengol Urp{\'\i}, Maria R Cervera, and Miguel AL Nicolelis. Analysis of neuronal ensemble activity reveals the pitfalls and shortcomings of rotation dynamics. \emph{Scientific Reports,} 9(1):1--14, 2019.
\bibitem{Michaels2016}Jonathan A Michaels, Benjamin Dann, and Hansj{\"o}rg Scherberger. Neural population dynamics during reaching are better explained by a dynamical system than representational tuning. \emph{PLoS computational biology,} 12(11):e1005175, 2016.
\bibitem{Fyhn2008}Marianne Fyhn, Torkel Hafting, Menno P Witter, Edvard I Moser, and May-Britt Moser. Grid cells in mice. \emph{Hippocampus,} 18(12):1230--1238, 2008.
\bibitem{Doeller2010}Christian F Doeller, Caswell Barry, and Neil Burgess. Evidence for grid cells in a human memory network. \emph{Nature,} 463(7281):657, 2010.
\bibitem{Yartsev2011}Michael M Yartsev, Menno P Witter, and Nachum Ulanovsky. Grid cells without theta oscillations in the entorhinal cortex of bats. \emph{Nature,} 479(7371):103, 2011.
\bibitem{Killian2012}Nathaniel J Killian, Michael J Jutras, and Elizabeth A Buffalo. A map of visual space in the primate entorhinal cortex. \emph{Nature,} 491(7426):761, 2012.
\bibitem{Jacobs2013}Joshua Jacobs, Christoph T Weidemann, Jonathan F Miller, Alec Solway, John F Burke, Xue-Xin Wei, Nanthia Suthana, Michael R Sperling, Ashwini D Sharan, Itzhak Fried, et al. Direct recordings of grid-like neuronal activity in human spatial navigation. \emph{Nature neuroscience,} 16(9):1188, 2013.
\bibitem{Constantinescu2016}Alexandra O Constantinescu, Jill X O'Reilly, and Timothy EJ Behrens. Organizing conceptual knowledge in humans with a gridlike code. \emph{Science,} 352(6292):1464--1468, 2016.
\bibitem{Bellmund2018}Jacob LS Bellmund, Peter G{\"a}rdenfors, Edvard I Moser, and Christian F Doeller. Navigating cognition: Spatial codes for human thinking. \emph{Science,} 362(6415):eaat6766, 2018.
\bibitem{Stensola2012}Hanne Stensola, Tor Stensola, Trygve Solstad, Kristian Fr{\o}land, May-Britt Moser, and Edvard I Moser. The entorhinal grid map is discretized. \emph{Nature,} 492(7427):72, 2012.
\bibitem{Mcnaughton2006}Bruce L McNaughton, Francesco P Battaglia, Ole Jensen, Edvard I Moser, and May-Britt Moser. Path integration and the neural basis of the 'cognitive map'. \emph{Nature Reviews Neuroscience,} 7(8):663--678, 2006.
\bibitem{devalois1990}Russell L DeValois and Karen K DeValois. \emph{Spatial vision,} volume 14. Oxford university press, 1990.
\bibitem{Nauhaus2012}Ian Nauhaus, Kristina J Nielsen, Anita A Disney, and Edward M Callahan. Orthogonal micro-organization of orientation and spatial frequency in primate primary visual cortex. \emph{Nature neuroscience,} 15(12):1683, 2012.
\bibitem{Sclar1982}G Sclar and RD Freedman. Orientation selectivity in the cat's striate cortex is invariant with stimulus contrast. \emph{Experimental brain research,} 46(3):457--461, 1982.
\bibitem{Morrone1982}M Concetta Morrone, DC Burr, and Lamberto Maffei. Functional implications of cross-orientation inhibition of cortical visual cells. I. Neurophysiological evidence. \emph{Proceedings of the Royal Society of London. Series B. Biological Sciences,} 216(1204):335--354,1982.
\bibitem{Albrecht1991}Duane G Albrecht and Wilson S Geisler. Motion selectivity and the contrast-response function of simple cells in the visual cortex. \emph{Visual neuroscience,} 7(6):531--546, 1991.
\bibitem{Heeger1992}David J Heeger. Normalization of cell responses in cat striate cortex. \emph{Visual neuroscience,} 9(2):181--197, 1992.
\bibitem{Deangelis1992a}GC DeAngelis, JG Robson, I Ohzawa, and RD Freedman. Organization of suppression in receptive fields of neurons in cat visual cortex. \emph{Journal of Neurophysiology,} 68(1):144--163, 1992.
\bibitem{Sillito1996}AM Sillito and HE Jones. Context-dependent interactions and visual processing in V1. \emph{Journal of Physiology-Paris,} 90(3-4):205--209, 1996.
\bibitem{Carandini1998}Matteo Carandini, J Anthony Movshon, and David Ferster. Pattern adaptation and cross-orientation interactions in the primary visual cortex. \emph{Neuropharmacology,} 37(4-5):501--511, 1998.
\bibitem{Carandini2012}Matteo Carandini and David J Heeger. Normalization as a canonical neural computation. \emph{Nature Reviews Neuroscience,} 13(1):51, 2012.
\bibitem{Ringach2010}Dario L Ringach. Population coding under normalization. \emph{Vision research,} 50(22):2223--2232, 2010.
\bibitem{Tring2018}Elaine Tring and Dario L Ringach. On the subspace invariance of population responses. \emph{Neurons, Behavior, Data Analysis, and Theory,} 2018.
\bibitem{Averbeck2006b}Bruno B Averbeck and Daeyeol Lee. Effects of noise correlations on information encoding and decoding. \emph{Journal of neurophysiology,} 95(6):3633--3644, 2006.
\bibitem{Hyvarinen2009}Aapo Hyv{\"a}rinen, Jarmo Hurri, and Patrick O Hoyer. \emph{Natural image statistics: A probabilistic approach to early computational vision,} volume 39. Springer Science \& Business Media, 2009.
\bibitem{Rumyantsev2020}Oleg I Rumyantsev, J{\'e}r{\^o}me A Lecoq, Oscar Hernandez, Yanping Zhang, Joan Savall, Rados{\l}aw Chrapkiewicz, Jane Li, Hongkui Zeng, Surya Ganguli, and Mark J Schnitzer. Fundamental bounds on the fidelity of sensory cortical coding. \emph{Nature,} 580(7801):100--105, 2020.
\bibitem{Wimmer2014}Klaus Wimmer, Duane Q Nykamp, Christos Constantinidis, and Albert Compte. Bump attractor dynamics in prefrontal cortex explains behavioral precision in spatial working memory. \emph{Nature neuroscience,} 17(3):431--439, 2014.
\bibitem{Bartolo2020}Ramon Bartolo, Richard C Saunders, Andrew R Mitz, and Bruno B Averbeck. Information-Limiting Correlations in Large Neural Populations. \emph{Journal of Neuroscience,} 40(8):1668--1678, 2020.
\bibitem{Laughlin1998}Simon B Laughlin, Rob R de Ruyter van Steveninck, and John C Anderson. The metabolic cost of neural information. \emph{Nature neuroscience,} 1(1):36, 1998.
\bibitem{cueva2020}Christopher J Cueva, Peter Y Wang, Matthew Chin, and Xue-Xin Wei. Emergence of functional and structural properties of the head direction system by optimization of recurrent neural networks. \emph{International Conference on Learning Representations (ICLR),} 2020.
\bibitem{Tenenbaum2000}Joshua B Tenenbaum, Vin De Silva, and John C Langford. A global geometric framework for nonlinear dimensionality reduction. \emph{Science,} 290(5500):2319--2323, 2000.
\bibitem{Roweis2000}Sam T Roweis and Lawrence K Saul. Nonlinear dimensionality reduction by locally linear embedding. \emph{Science,} 290(5500):2323--2326, 2000.
\bibitem{Chapin1999}John K Chapin and Miguel AL Nicolelis. Principal component analysis of neuronal ensemble activity reveals multidimensional somatosensory representations. \emph{Journal of neuroscience methods,} 94(1):121--140, 1999.
\bibitem{Briggman2005}Kevin L Briggman, Henry DI Abarbanel, and William B Kristan. Optical imaging of neuronal populations during decision-making. \emph{Science,} 307(5711):896--901, 2005.
\bibitem{Macke2011}Jakob H Macke, Lars Buesing, John P Cunningham, M Yu Byron, Krishna V Shenoy, and Maneesh Sahani. Empirical models of spiking in neural populations. In \emph{Advances in neural information processing systems,} pages 1350--1358, 2011.
\bibitem{Mante2013}Valerio Mante, David Sussillo, Krishna V Shenoy, and William T Newsome. Context-dependent computation by recurrent dynamics in prefrontal cortex. \emph{Nature,} 503(7474):78--84, 2013.
\bibitem{Sadtler2014}Patrick T Sadtler, Kristin M Quick, Matthew D Golub, Steven M Chase, Stephen I Ryu, Elizabeth C Tyler-Kabara, Byron M Yu, and Aaron P Batista. Neural constraints on learning. \emph{Nature,} 512(7515):423--426, 2014.
\bibitem{Pandarinath2018}Chethan Pandarinath, Daniel J O'Shea, Jasmine Collins, Rafal Jozefowicz, Sergey D Stavisky, Jonathan C Kao, Eric M Trautmann, Matthew T Kaufman, Stephen I Ryu, Leigh R Hochberg, et al. Inferring single-trial neural population dynamics using sequential auto-encoders. \emph{Nature methods,} 15(10):805--815, 2018.
\bibitem{Minxha2020}Juri Minxha, Ralph Adolphs, Stefano Fusi, Adam N Mamelak, and Ueli Rutishauser. Flexible recruitment of memory-based choice representations by the human medial frontal cortex. \emph{Science,} 368(6498), 2020.
\bibitem{Singh2008}Gurjeet Singh, Facundo Memoli, Tigran Ishkhanov, Guillermo Sapiro, Gunnar Carlsson, and Dario L Ringach. Topological analysis of population activity in visual cortex. \emph{Journal of vision,} 8(8):11, 2008.
\bibitem{Giusti2015} Chad Giusti, Eva Pastalkova, Carina Curto, and Vladimir Itskov. Clique topology reveals intrinsic geometric structure in neural correlations. \emph{Proceedings of the National Academy of Sciences}, 112(44):13455--13460, 2015.
\bibitem{Chaudhuri2019}Rishidev Chaudhuri, Berk Gercek, Biraj Pandey, Adrien Peyrache, and Ila Fiete. The intrinsic attractor manifold and population dynamics of a canonical cognitive circuit across waking and sleep. \emph{Nature neuroscience,} 22(9):1512--1520, 2019.
\bibitem{Low2018}Ryan J Low, Sam Lewallen, Dmitriy Aronov, Rhino Nevers, and David W Tank. Probing variability in a cognitive map using manifold inference from neural dynamics. \emph{bioRxiv,} page 418939, 2018.
\bibitem{Zhou2020} Ding Zhou, and Xue-Xin Wei. Learning identifiable and interpretable latent models of high-dimensional neural activity using pi-VAE. \emph{Advances in Neural Information Processing Systems (NeurIPS)}, 2020.
\bibitem{Langer2001}Michael S Langer and Heinrich H B{\"u}lthoff. A prior for global convexity in local shape-from-shading. \emph{Perception,} 30(4):403--410, 2001.
\bibitem{Weiss2002}Yair Weiss, Eero P Simoncelli, and Edward H Adelson. Motion illusions as optimal percepts. \emph{Nature neuroscience,} 5(6):598--604, 2002.
\bibitem{Adams2004}Wendy J Adams, Erich W Graf, and Marc O Ernst. Experience can change the 'light-from-above' prior. \emph{Nature neuroscience,} 7(10):1057--1058, 2004.
\bibitem{Griffiths2006}Thomas L Griffiths and Joshua B Tenenbaum. Optimal predictions in everyday cognition. \emph{Psychological science,} 17(9):767--773, 2006.
\bibitem{Stocker2006}Alan A Stocker and Eero P Simoncelli. Noise characteristics and prior expectations in human visual speed perception. \emph{Nature neuroscience,} 9(4):578--585, 2006.
\bibitem{Kording2006}Konrad P K{\"o}rding and Daniel M Wolpert. Bayesian decision theory in sensorimotor control. \emph{Trends in cognitive sciences,} 10(7):319--326, 2006.
\bibitem{Gibson1979}James J Gibson. The Ecological Approach to Visual Perception. 1979.
\bibitem{Shepard1984}Roger N Shepard. Ecological constraints on internal representation: resonant kinematics of perceiving, imagining, thinking, and dreaming. \emph{Psychological review,} 91(4):417, 1984.
\bibitem{Shepard2001}Roger N Shepard. Perceptual-cognitive universals as reflections of the world. \emph{Behavioral and brain sciences,} 24(4):581--601, 2001.
\bibitem{Wiskott2002}Laurenz Wiskott and Terrence J Sejnowski. Slow feature analysis: Unsupervised learning of invariances. \emph{Neural computation,} 14(4):715--770, 2002.
\bibitem{Berkes2005}Pietro Berkes and Laurenz Wiskott. Slow feature analysis yields a rich repertoire of complex cell properties. \emph{Journal of vision,} 5(6):9, 2005.
\bibitem{Gardenfors2004}Peter G{\"a}rdenfors. \emph{Conceptual spaces: The geometry of thought.} MIT press, 2004.
\bibitem{Fechner1860}Gustav Theodor Fechner. \emph{Elemente der psychophysik,} volume 2. Breitkopf u. H{\"a}rtel, 1860.
\bibitem{Thurstone1927}Louis L Thurstone. A law of comparative judgment. \emph{Psychological review,} 34(4):273, 1927.
\bibitem{Attneave1950}Fred Attneave. Dimensions of similarity. \emph{The American journal of psychology,} 63(4):516--556, 1950.
\bibitem{Ekman1954}Gosta Ekman. Dimensions of color vision. \emph{The Journal of Psychology,} 38(2):467--474, 1954.
\bibitem{Beals1968}Richard Beals, David H Krantz, and Amos Tversky. Foundations of multidimensional scaling. \emph{Psychological review,} 75(2):127, 1968.
\bibitem{Shepard1964b}Roger N Shepard. Attention and the metric structure of the stimulus space. \emph{Journal of mathematical psychology,} 1(1):54--87, 1964.
\bibitem{Krantz1975}David H Krantz and Amos Tversky. Similarity of rectangles: An analysis of subjective dimensions. \emph{Journal of Mathematical Psychology,} 12(1):4--34, 1975.
\bibitem{Shepard1987}Roger N Shepard. Toward a universal law of generalization for psychological science. \emph{Science,} 237(4820):1317--1323, 1987.
\bibitem{Nosofsky1986}Robert M Nosofsky. Attention, similarity, and the identification--categorization relationship. \emph{Journal of experimental psychology: General,} 115(1):39, 1986.
\bibitem{Shepard1962}Roger N Shepard. The analysis of proximities: multidimensional scaling with an unknown distance function. \emph{Psychometrika,} 27(2):125--140, 1962.
\bibitem{Torgerson1965}Warren S Torgerson. Multidimensional scaling of similarity. \emph{Psychometrika,} 30(4):379--393, 1965.
\bibitem{Tversky1977}Amos Tversky. Features of similarity. \emph{Psychological review,} 84(4):327, 1977.
\bibitem{Shepard1980}Roger N Shepard. Multidimensional scaling, tree-fitting, and clustering. \emph{Science,} 210(4468):390--398, 1980.
\bibitem{Tversky1982}Amos Tversky and Itamar Gati. Similarity, separability, and the triangle inequality. \emph{Psychological review,} 89(2):123, 1982.
\bibitem{Ashby1988}F Gregory Ashby and Nancy A Perrin. Toward a unified theory of similarity and recognition. \emph{Psychological review,} 95(1):124, 1988.
\bibitem{Kriegeskorte2012a}Nikolaus Kriegeskorte and Marieke Mur. Inverse MDS: Inferring dissimilarity structure from multiple item arrangements. \emph{Frontiers in psychology,} 3:245, 2012.
\bibitem{Hebart2020}Martin Hebart, Charles Y Zheng, Francisco Pereira, and Chris Baker. Revealing the multidimensional mental representations of natural objects underlying human similarity judgments. 2020.
\bibitem{Mur2013}Marieke Mur, Mirjam Meys, Jerzy Bodurka, Rainer Goebel, Peter A Bandettini, and Nikolaus Kriegeskorte. Human object-similarity judgments reflect and transcend the primate-IT object representation. \emph{Frontiers in psychology,} 4:128, 2013.
\bibitem{Cichy2019}Radoslaw M Cichy, Nikolaus Kriegeskorte, Kamila M Jozwik, Jasper JF van den Bosch, and Ian Charest. The spatiotemporal neural dynamics underlying perceived similarity for real-world objects. \emph{Neuroimage,} 194:12--24, 2019.
\bibitem{Golden2016}James R Golden, Kedarnath P Vilankar, Michael CK Wu, and David J Field. Conjectures regarding the nonlinear geometry of visual neurons. \emph{Vision research,} 120:74--92, 2016.
\bibitem{Chung2018}SueYeon Chung, Daniel D Lee, and Haim Sompolinsky. Classification and geometry of general perceptual manifolds. \emph{Physical Review X,} 8(3):031003, 2018.
\bibitem{Sohn2019}Hansem Sohn, Devika Narain, Nicholas Meirhaeghe, and Mehrdad Jazayeri. Bayesian computation through cortical latent dynamics. \emph{Neuron,} 103(5):934--947, 2019.
\bibitem{Henaff2019}Olivier J H{\'e}naff, Robbe LT Goris, and Eero P Simoncelli. Perceptual straightening of natural videos. \emph{Nature neuroscience,} 22(6):984--991, 2019.
\bibitem{Kafashan2021} MohammadMehdi Kafashan, Anna W Jaffe, Selmaan N Chettih, Ramon Nogueira, I{\~n}igo Arandia-Romero, Christopher D Harvey, Rub{\'e}n Moreno-Bote, and Jan Drugowitsch. Scaling of sensory information in large neural populations shows signatures of information-limiting correlations. \emph{Nature Communications,} 12(1):1--16, 2021.
\bibitem{Nogueira2020} Ramon Nogueira, Nicole E Peltier, Akiyuki Anzai, Gregory C DeAngelis, Julio Mart{\'\i}nez-Trujill, and Rub{\'e}n Moreno-Bote. The effects of population tuning and trial-by-trial variability on information encoding and behavior. \emph{Journal of Neuroscience,} 40(5):1066--1083, 2020.
\bibitem{Okazawa2021} Gouki Okazawa, Christina E Hatch, Allan Mancoo, Christian K Machens, and Roozbeh Kiani. The geometry of the representation of decision variable and stimulus difficulty in the parietal cortex. \emph{bioRxiv}, 2021. 
\bibitem{Amari2000} Shun-ichi Amari. Information geometry.
\emph{Japanese Journal of Mathematics}, 1--48, 2000.
\bibitem{Bhattacharyya1943} Anil Bhattacharyya. On a measure of divergence between two statistical populations defined by their probability distributions. \emph{Bull. Calcutta Math. Soc.} 35:99--109, 1943
\bibitem{Johnson2001} Don Johnson, and Sinan Sinanovic. Symmetrizing the Kullback-Leibler distance. \emph{IEEE Transactions on Information Theory}, 2001.

\end{thebibliography}
\end{document}